\documentclass[epsfig,12pt]{article}
\usepackage{epsfig}
\usepackage{graphicx}
\usepackage{array}
\usepackage{color}
\usepackage{bm}
\usepackage{cite}
\usepackage{geometry}
\usepackage{latexsym}
 \usepackage{amsmath, amssymb, amscd, xypic, graphicx}

\newcommand{\beq}{\begin{equation}}   
\newcommand{\eeq}{\end{equation}}
\newcommand{\beqn}{\begin{eqnarray}}   
\newcommand{\eeqn}{\end{eqnarray}}

\def\ntwo{${\mathcal N}=2\;$}

\def\ntwot{${\mathcal N}=(2,2)\;$}

\newcommand*\xbar[1]{%
 \kern0.5ex%
  \hbox{%
   \kern0.2ex%
      \vbox{%
      \hrule height 0.5pt % The actual bar
      \kern0.5ex%         % Distance between bar and symbol
      \hbox{%
        \kern-0.1em%      % Shortening on the left side
        \ensuremath{#1}%
        \kern-0.1em%      % Shortening on the right side
      }%
    }%
  }%
}

\newcommand{\gsim}{\lower.7ex\hbox{$
\;\stackrel{\textstyle>}{\sim}\;$}}
\newcommand{\lsim}{\lower.7ex\hbox{$
\;\stackrel{\textstyle<}{\sim}\;$}}
\begin{document}

\begin{titlepage}

\begin{flushright}
FTPI-MINN-18/06, UMN-TH-3715/18
\end{flushright}

\vspace{5mm}

\begin{center}
{  \Large \bf  
Supersymmetric Tools in Yang-Mills Theories at\\[2mm]
 Strong Coupling: the Beginning of a Long \\[2mm]Journey\,\footnote{Based on the talk at the Dirac Medal Award Ceremony, ICTP, March 22, 2017.}}
 
 \vspace{6mm}

{\Large Mikhail Shifman} 

\vspace{3mm}

{\it  William I. Fine Theoretical Physics Institute,
University of Minnesota,
Minneapolis, MN 55455, USA}
\end{center}

\vspace{4mm}

\vspace{6mm}

\begin{center}
{\large\bf Abstract}
\end{center}

Development of holomorphy-based methods in super-Yang-Mills theories started in the early 1980s
and lead to a number of breakthrough results. I review some results in which I participated.
The discovery of Seiberg's duality and the Seiberg-Witten solution of ${\cal N}=2$ Yang-Mills were the milestones
in the long journey of which, I assume, much will be said in other talks. I will focus on the discovery (2003) 
of non-Abelian vortex strings with various degree of supersymmetry, supported in some four-dimensional Yang-Mills theories 
and some intriguing implications of this discovery. One of the recent results is the observation of a  soliton string in the bulk ${\cal N}=2$ theory with the $U(2)$ gauge group and four flavors,  which can become critical in a certain limit. This is the case of
a ``reverse holography,'' with a very transparent physical meaning.

\vspace{2cm}

\end{titlepage}

\begin{quote}

{\small \em I should say that receiving the Dirac Medal is the highest honor for me as a theoretical physicist. I am deeply 
grateful to Abdus Salam Center for Theoretical Physics for awarding me this highly prestigious Prize. }

\end{quote}

\section{Introduction}
\label{intro}

So far quantum field theory (QFT) remains the basis of our understanding of fundamental laws of Nature. QFT is ninety years old. If you look at books and reviews devoted to QFT written in the 1950s and '60s and compare them with today's reviews, you will hardly say this is one and the same discipline. Many questions on which we focus at present could not even be formulated then. 

Since then QFT underwent two dramatical changes: the discovery and development of Yang-Mills theories which, as we know, run our four-dimensional world, and the discovery and development of supersymmetric gauge theories which dominate the modern theoretical landscape.  The first {\em four}-dimensional supersymmetric theory -- supergeneralization of 
quantum electrodynamics -- was constructed in 1970 and published in \cite{GL} by Golfand and Likhtman. The breakthrough works of Wess and Zumino paved the way to remarkable advances in this area which continue unabated forty years later.

\begin{figure}[h]
\centerline{\includegraphics[width=11cm]{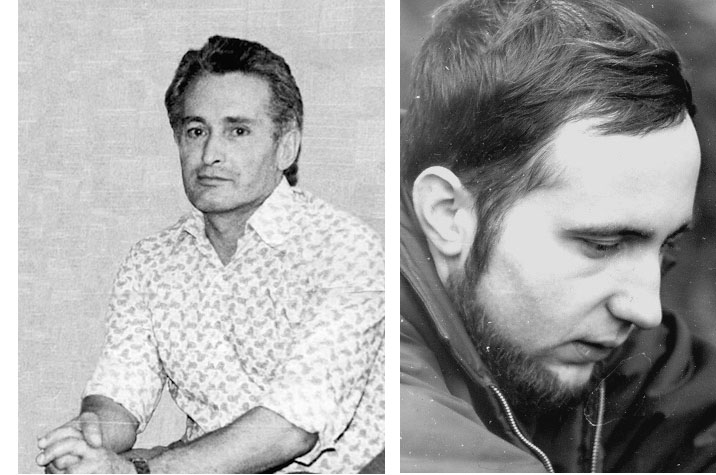}}
\caption{\small Yury Golfand and Evgeny Likhtman, circa 1980.}
\label{figmeson} 
\end{figure}

Although none of the expected superpartners have been experimentally detected so far supersymmetry changed  
quantum field theory at a deep level by providing answers to some hard questions, such as divergences in the vacuum energy density. Any theory which can be embedded in ${\cal N}=4$ super-Yang-Mills is ultraviolet finite. 

The first exact statement of this type was the vanishing of the vacuum energy in supersymmetric theories.\footnote{As early as in 1950s,  Wolfgang Pauli delivered a landmark series of lectures at the Swiss Federal Institute of Technology (ETH) in Zurich. They were published in English by MIT Press only in 1973.
In Section 9 of Volume 6 \cite{PauliLec} Pauli discusses the vacuum energy density in various field theories
known at that time. He observes that adding the Dirac spinor contribution to that of two complex scalar fields cancels divergences and produces zero vacuum energy -- the first ever hint to supersymmetry!  
In the context of supersymmetric field theories the zero vacuum energy was first observed in \cite{zve}.
} Endowing Yang-Mills theories with supersymmetry creates a broad class of the so-called protected quantities which can be calculated exactly in both weak and strong coupling regimes.
In 1982-83 we observed \cite{NSVZinst} that not only in the vacuum, but in {\em all} backgrounds which conserve a part of supercharges exact results are attainable. Analysis of instantons \cite{BPST} in super-Yang-Mills theories \cite{NSVZinst,NSVZc} was the beginning of a long journey on which we embarked and of which I am going to speak today. I will only briefly  summarize ``old" results
since they were reviewed in detail in my 1999 Sakurai talk \cite{sak}. Then I will pass to a new era which to my mind is associated with  the breakthrough solution of slightly deformed ${\cal N}=2$ super-Yang-Mills theory by Seiberg and Witten \cite{sei}.

\section{Setting the stage (1980s) and later}

\subsection{Instantons and \boldmath{$\beta$} functions in four and two dimensions}

The instanton background field conserves half of the supercharges. The other half acts on instanton nontrivially producing fermion zero modes. The number of the instanton
zero modes   is related to the number of nontrivially
realized symmetries. 
The supercharges that are conserved guarantee that all quantum corrections cancel, and hence the instanton measure can be
exactly calculated, impying, in turn, the exact $\beta$ functions. In {\em pure} super-Yang-Mills theories with various degree of supersymmetry we obtain
\beq
\beta (\alpha) = - \left(n_b-\frac{n_f}{2}\right)\, \frac{ 
\alpha^2}{2\pi}\left[
1-\frac{\left(n_b-n_f\right)\,\alpha}{4\pi}\right]^{-1}\, ,
\label{totbetapg3}
\eeq
where $n_b$ and $n_f$ count  the 
gluon and gluino zero modes, respectively. For ${\cal N}= 1$ we have $n_b=2n_f = 4T_G$ where
$T_G$ is the dual Coxeter number (it is also called
 1/2 the Dynkin index; for $SU(N)$ we have $T_G=N$).
 For ${\cal N}= 2$,
one gets $n_b = n_f = 4T_G$, implying that the $\beta$ function reduces to 
one-loop. For ${\cal N}= 4$ the $\beta$ function vanishes since  $n_f = 2n_b$.

The main lesson obtained in \cite{NSVZinst} was as follows. Equation (\ref{totbetapg3}) makes explicit that {\em all} coefficients
of the $\beta$ functions in {\em pure} super-Yang-Mills theories  have a geometric origin since they are in one-to-one correspondence with the number of symmetries nontrivially realized on instantons.

In  theories with matter, apart from the gluon 
and  gluino zero modes, one has to deal with the zero 
modes of the matter 
fermions. While the gluon/gluino $Z$ factors are related to the 
gauge coupling constant $g^2$ itself, this is not the case for the
$Z$ factors of the matter fermions.  Therefore, in  theories with
matter the exact  instanton measure implies an exact relation
between the $\beta$ function and the anomalous dimensions 
$\gamma_i$,
\begin{equation}
\beta (\alpha) = -\frac{\alpha^2}{2\pi}\left[3\,T_G -\sum_i T(R_i)(1-
\gamma_i )
\right]\left(1-\frac{T_G\,\alpha}{2\pi} \right)^{-1}
\, ,
\label{nsvzbetaf}
\end{equation}
where $T(R_i)$ is the Dynkin index in the representation $R_i$,
$$
\mbox{Tr}\, (T^a T^b ) = T(R_i)\, \delta^{ab}\, ,
$$
and $T^a$ stands for the generator of the gauge group $G$ in the appropriate representation
which can be arbitrary.

Equation (\ref{nsvzbetaf})
is valid for
arbitrary Yukawa interactions of  the matter fields. The Yukawa 
interactions show
up only through the  anomalous dimensions. It appeared first in \cite{nsvzbetamatter}, and shortly
after in a more closed form in \cite{ShifmanVainshtein}.\footnote{For the current status of direct perturbative derivations 
of the NSVZ $\beta$ function see
\cite{Step} and references therein. } The latter work presented the solution of the
anomaly supermultiplet problem by virtue of the Wilsonean approach, more of which will be said in Arkady Vainshtein's talk. 
Recently the anomaly supermultiplet problem was revisited  \cite{Ko} and extended to a number of theories
which were not considered in the 1980s.  

Equation (\ref{nsvzbetaf}) played a crucial role in establishing the edges of the conformal window
in Seiberg's duality \cite{Seiberg}.

By the same token exact $\beta$ functions can be obtained and the geometric nature of the coefficients revealed
 in {\em two-dimensional} sigma models. In the 1980s this was done for the ${\cal N}= (2,2)$ sigma models \cite{sigma}, i.e. 
 the model with extended supersymmetry. Recently an interesting class of ``heterotic" ${\cal N}= (0,2)$ sigma models was discovered in connection with non-Abelian strings which will be discussed later. It turns out that the chiral ${\cal N}= (0,2)$ sigma models
 (in which there is no symmetry between the left- and right-moving fermions) represent a direct analog 
 of four-dimensional Yang-Mills.  This was an exciting finding.
 
 Thus, in the minimal heterotic $CP(1)$ model, in which all left-handed fermions are dropped\,\footnote{The minimal models of this type 
 do not exist for $CP(N-1)$ with $N>2$ because of the anomaly \cite{Moore,Chen}.}  we have  \cite{cui1}
\beq
\beta (g^2) = -\frac{g^4}{4\pi}\,\frac{1}{1-\frac{g^2}{4\pi}}\,,
\label{sigma3}
\eeq 
 which is a direct analogue of Eq. (\ref{totbetapg3}) in super-Yang-Mills. In this particular $CP(1)$ model with 
 ${\cal N}= (0,2)$ supersymmetry one can introduce  ``matter," i.e. $N_f$ superfields of ${\cal N}= (0,2)$ supersymmetry
which contain only a single left-moving fermion degree of freedom (per flavor). The fermion state in this multiplet
has no bosonic counterpart -- other superfield components are auxiliary.

In this model, instead of (\ref{sigma3}) we arrive at \cite{cui1}
\beq
\beta (g^2) = -\frac{g^4}{4\pi}\,\frac{1+\frac{N_f\,\gamma}{2}}{1-\frac{g^2}{4\pi}}\,,
\label{sigma4}
\eeq 
where $\gamma$ is the anomalous dimension of the ``matter" superfield. This expression is similar to that
in Eq. (\ref{nsvzbetaf}). The only difference is that the ``matter" field fermions do not show up in the numerator at {\em one} loop; they appear only
in the second and higher loops.
This is a special feature of the two-dimensional $CP(N-1)$ models. 

If $N_f=1$ then ${\cal N}=(0,2)$ is uplifted to (2,2) and the $\beta$ function in (\ref{sigma4}) becomes one-loop. 
The numerator in  (\ref{sigma4}) is exactly canceled by the denominator. At sufficiently large $N_f$ 
Eq. (\ref{sigma4}) obviously exhibits an infrared fixed point. This is similar to the Banks-Zaks phenomenon \cite{Banks}. Therefore, in  the minimal heterotic (0,2) $CP(1)$ model a conformal
window exists. The question of whether a dual representation exists in this model is still open. I think this is a very challenging question.

The above results were extended to $CP(N-1)$  with $N>2$ \cite{cui2}. To this end the minimal model could not be used, as was mentioned above: a quantum anomaly makes it inconsistent. However, the {\em non}-minimal ${\cal N}= (0,2)$ model discovered  in the studies of non-Abelian vortex-strings supported in ${\cal N}=1$ super-Young-Mills \cite{shiy} is selfconsistent for any 
$N$. The non-minimal model is obtained from the conventional $(2,2)$ model by deforming it by an extra right-moving field 
$\zeta$ with a heterotic coupling parametrized by a constant $h$. In this case we derived \cite{cui2} the exact relation between the $\beta$ function 
and the anomalous dimensions $\gamma$ of the ``matter" fields,
\beq
\beta_{g}\!=\!-\frac{g^{2}}{4\pi}\,\,\frac{N\,g^{2}\left(1+\gamma_{\psi_R}/2
\right) - {h}^{2}\left(\gamma_{\psi_{R}}+\gamma_{\zeta}\right)}{1-({h}^{2}/4\pi)}\,,
\label{sigma5}
\eeq
where $\gamma_{\psi_{R}},\,\gamma_{\zeta}$ are the anomalous dimensions of the $\psi_{R},\,\zeta_{R}$ fields, respectively.
Here $g^2$ is the constant parametrizing the $CP(N-1)$ target space, $h$ is a heterotic deformation parameter, and  $\zeta_{R}$ is an additional field  breaking (2,2)
down to (0,2).  It has no left-moving counterpart. Equation (\ref{sigma5}) has the same structure as that in ${\cal N}=1$ Yang-Mills with matter, see (\ref{nsvzbetaf}).

\subsection{Adler functions}

\begin{figure}[h]
\centerline{\includegraphics[width=7cm]{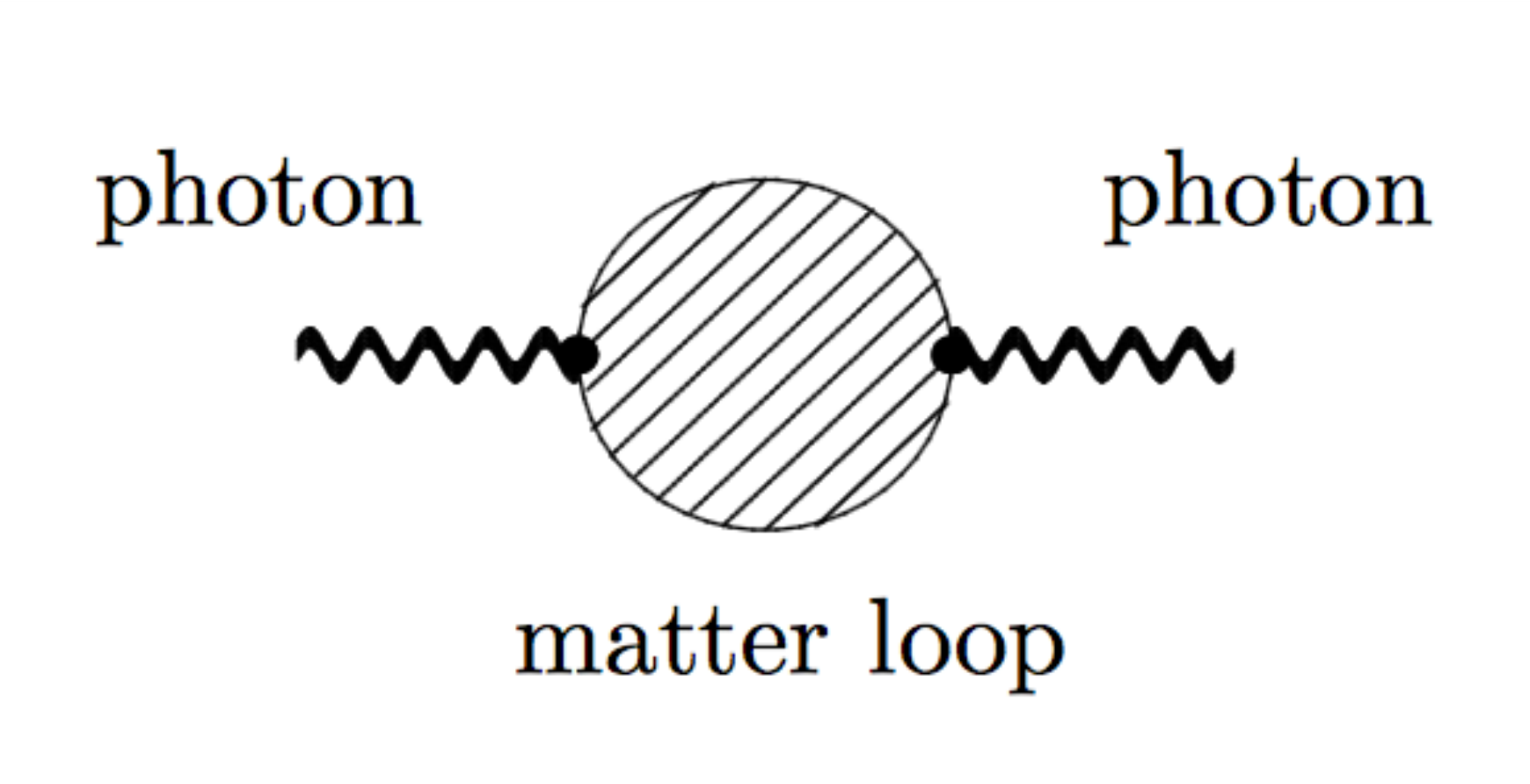}}
\caption{\small  The Adler function in ${\cal N}=1$ SQCD. The shaded circle presents all loop graphs. Borrowed from the second paper in \cite{Step}.}
\label{af} 
\end{figure}

The above ideas that allowed us to obtain the exact $\beta$ functions could have been used in the 1980s to obtained exact predictions for the
Adler $D$ functions in SQCD. This never happened, however. Only in 2015 this gap was filled \cite{SS}. The Adler  function is defined as an infinite set of the 
diagrams symbolically depicted in Fig. \ref{af}.

\begin{figure}[h]
\centerline{\includegraphics[width=12cm]{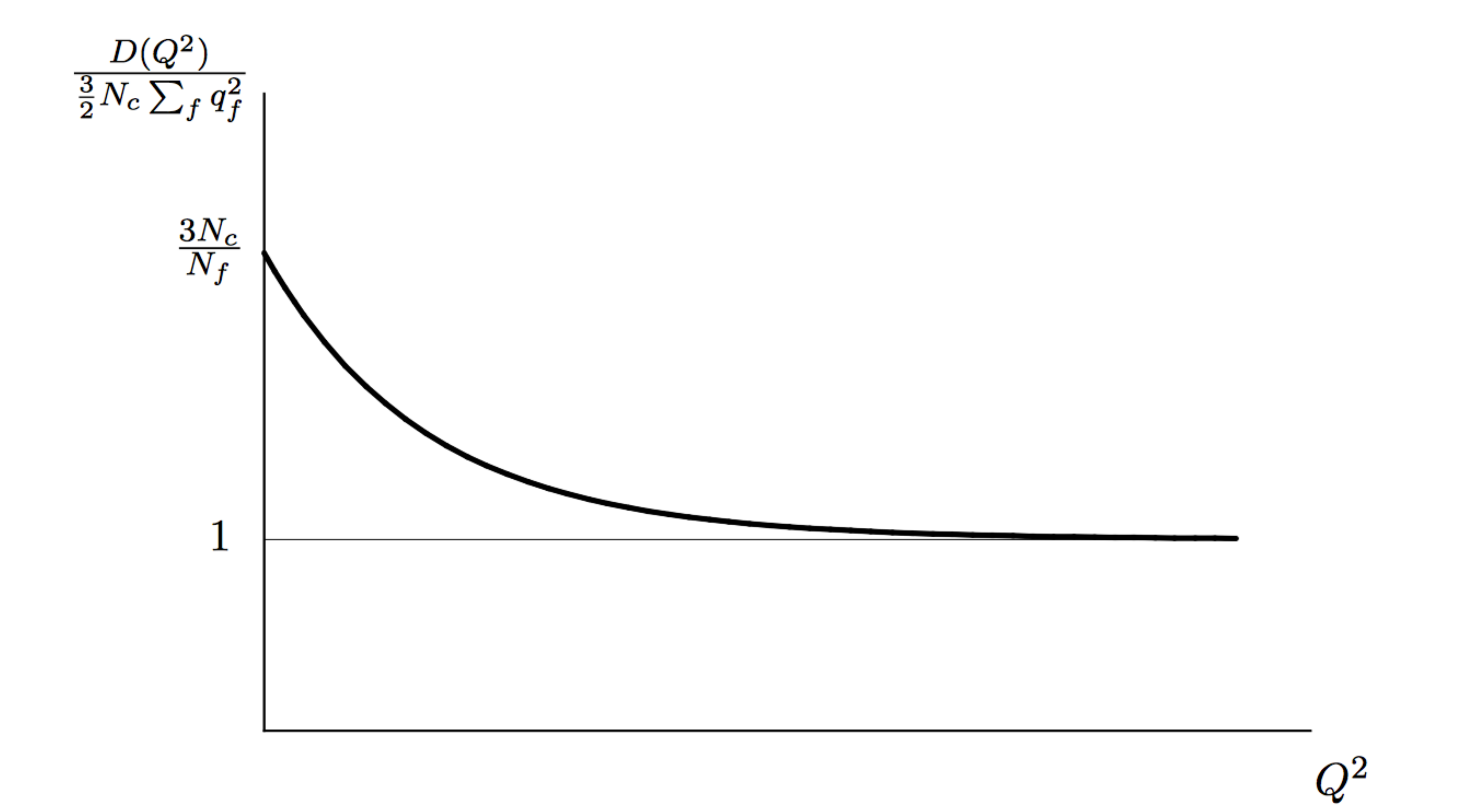}}
\caption{\small $ D(Q^2) \left(\frac 32 N_c \sum_f q_f^2 \right)^{-1}$ versus $Q^2$. The horizontal lines corresponds to $N_f = 3N_c$, i.e. the right edge of the conformal window.}
\label{window} 
\end{figure}

\noindent
The exact result is
\beq
D(Q^2) = \frac 32 N_c \sum_f q_f^2 \Big[1- \gamma (\alpha_s (Q^2))\Big]
\label{tsix}
\eeq
where the sum runs over all flavors, $q_f$ is the electric charge of a given flavor, $\gamma$ is the anomalous dimension of the matter
fields. It is the same for all matter fields assuming that they all belong to the fundamnental representation of color. Moreover, $\alpha_s$ is the strong coupling constant. The result is plotted in Seiberg's conformal window $\frac 32 N_c < N_f < 3N_c$
in Fig. \ref{window}.

For a very recent discussion of Eq. (\ref{tsix}) in perturbation theory I refer the reader to \cite{KKS}. This latter paper is based on a development in 
supersymmetric perturbation theory which I'd like to mention \cite{KS}. It is known under the name the ``NSVZ regularization scheme,"  with 
 higher derivatives. We anticipated  the existence of such a scheme  in the 1980s. It took 30 years to develop it and make it work!

\newpage

\subsection{Gluino condensate and localization}
\label{22}

Simultaneously with the first exact $\beta$ functions we found a way to calculate
the gluino condensate \cite{gc}
\beq
{\cal O}_\lambda = {\rm Tr} \,\left( \lambda^\alpha \lambda_\alpha\right).
\eeq
This problem tortured us for at least two years, until we realized that if one considers $n$-point functions of operators 
 which are $Q$-closed (such as ${\cal O}_\lambda$), while their spacial derivatives are $Q$-exact\,\footnote{In  the case of the gluino condensate $\left[ {\bar Q}^{\dot\beta}\,, {\cal O}_\lambda\right] = 0$ while $\partial_{\alpha\dot\beta}{\cal O}_\lambda \sim 
\left\{ {\bar Q}^{\dot\beta} \,, {\rm Tr}\, \left(\lambda^\beta G_{\beta\alpha}\right)\right\}$. }
 then the above  $n$-point functions must be independent of coordinates.
For example, the $SU(2)$ instanton completely saturates the following two-point function:  \beq
\big\langle T \left[{\cal O}_\lambda (x) {\cal O}_\lambda (0)\right]\big\rangle = c_s^2 \, \Lambda^6\,, \quad \mbox{for}\,\, \forall\,\, x\,,
\label{lambda1}
\eeq
in ${\cal N}=1$
super-Yang-Mills theory with the $SU(2)$ gauge group. 
Here $c_s$ is an exactly calculable numerical constant and $\Lambda$ is the dynamical scale. The exact gluino condensate
follows from (\ref{lambda1}),
\beq
\big\langle {\cal O}_\lambda \big\rangle = \pm c_s \, \Lambda^3\,.
\label{lambda2}
\eeq
I think we were the first to prove the above theorem (please, correct me if I am wrong). By now it became instrumental in 
many supersymmetry-based constructions, for instance, in Nekrasov's localization, of which I will say a few words later.

The calculation of (\ref{lambda1})  was performed at short distances, $|x| \ll \Lambda^{-1}$, where this correlation function is saturated by instantons of small size, $\rho \sim | x |$. However, the theory itself -- ${\cal N}=1$
pure Yang-Mills -- is strongly coupled. Although the result in (\ref{lambda1}) was perfectly well-defined mathematically
I still could not get rid of a feeling of vague dissatisfaction. It was desirable to revisit the problem at weak coupling (in the Higgs regime),
calculate the gluino condensate enjoying full theoretical control over the theory, and then analytically continue
back to strong coupling using the known holomorphic behavior of the condensate on  the {\em mass} parameter $m$  of the Higgs field. On general grounds it was known that in the $SU(2)$ gauge theory 
$\big\langle {\cal O}_\lambda \big\rangle \sim \sqrt{m}$.
The only singularity is of the square root type -- this statement is exact. The simple analytic continuation method we developed in 1985 perhaps can be viewed as a primitive rudimentary precursor to the powerful Seiberg-Witten solution.

The result for the gluino condensate obtained in this way \cite{Novikov} (see also \cite{masv}) after taking $m$ to be equal to the ultraviolet cut-off has the same functional dependence as in (\ref{lambda2}), namely
$\big\langle {\cal O}_\lambda \big\rangle = \pm c_w \, \Lambda^3$ where the subscript $w$ means the weak coupling calculation
as opposed to $c_s$ appearing at strong coupling.  Originally it was found \cite{Novikov} that $c_s = \sqrt\frac 45\, c_w$. 
Now there is no doubt (see e.g. \cite{khoze}) that it is the weak coupling calculation that produces the correct value of the 
coefficient in the gluino condensate. A physical explanation of the discrepancy above is not found yet.

The paper \cite{Novikov} was entitled ``Supersymmetric Instanton Calculus (Gauge Theories with Matter)."
The theory we dealt with to determine the gluino condensate in the weak coupling (Higgs) regime was well-defined at all distances, including  large, 
because of complete Higgsing. Under these circumstances we were certain that $\big\langle {\cal O}_\lambda \big\rangle$ was saturated by one instanton. It was
natural to think that the size of the saturating instanton would be of the order of $\rho \sim v^{-1}$
where $v$ is the vacuum expectation value of the Higgs field. 
However, a great surprise expected us {\em en route}.
A remarkable observation  was made in \cite{Novikov}: 
the integral over the instanton size proved to be  {\em completely} saturated
by {\em zero-size} instantons, because we represented it as
 a full derivative which reduced to a delta function, $\delta (\rho^2)$ in the integrand.
 This was probably the first example of self-{\em localization} which later grew into a powerful 
construction of the Nekrasov localization \cite{Nekrasov}.

\subsection{Mass spectrum}

The first superalgebra  in four-dimensional 
field theory was derived by Golfand  and
Likhtman  \cite{GL} in the form
\begin{equation}
\{\bar Q_{\dot\alpha} Q_\beta\} = 2 
P_\mu \left( \sigma^\mu \right)_{\alpha\beta}\,,\qquad
\{\bar Q_\alpha \bar Q_\beta\}= \{  Q_\alpha Q_\beta\}=0\,,
\label{glsa}
\end{equation}
i.e. with no central extensions. Possible occurrence of extensions (elements of
superalgebra commuting with other generators) was first mentioned in an
unpublished paper of  \L{}opusza\'{n}ski\index{Lopuszanski} and 
Sohnius\index{Sohnius} \cite{A2}
where the last two anticommutators were modified as (see also \cite{A3})
\begin{equation}
\{  Q_\alpha^I  Q_\beta^G \} = Z_{\alpha \beta}^{IG}\,.
\end{equation}
The superscripts $I,G$ mark extended supersymmetry.
The central charge derived in this paper was that in ${\mathcal N}=2$
superalgebra in four dimensions,
$Z_{\alpha\beta}^{IG} \sim \varepsilon_{\alpha\beta}\varepsilon^{IG}$. It is 
Lorentz scalar and is relevant to the magnetic monopole (dyon) masses. 
In the 1980s the focus was on consequences from the
Lorentz scalar central charges.

It was generally understood that superalgebras with (Lorentz-scalar)
central charges can be obtained from superalgebras without
central charges in higher-dimensi\-onal space-time by interpreting some
of the extra components of the momentum as central charges. If the ``central charges'' carry Lorentz indices  
or contain contributions from quantum anomalies (or both), the dimension reduction
procedure cannot be used. Above I put ``central charges" in the quotation
marks because $Z_{\{\alpha \beta\}}$ or $Z_\mu$ or other Lorentz-noninvariant elements
of superalgebras in various dimensions
are {\em not} central in the strict sense of the word: they only commute with $Q_\alpha$, $\bar Q_{\dot\alpha}$
 and $P_\mu$, and do not commute with Lorentz rotations, since they carry the Lorentz indices.
 They are  associated with extended topological defects --- such as domain walls or strings ---  
and could be called {\em brane charges} \cite{Ko}. Leaving this subtlety aside, 
 I will continue to refer
 to these elements as {\em central charges} (CCs), or, sometimes, 
 tensorial central charges.
 
 Why the central extension of (\ref{glsa}) is so important? In such theories the so-called 
 short supermultiplets
 exist. Because they are short their mass (or tension in the case of extended objects)
 is exactly equal to the corresponding CC \cite{Olive}. The latter are of topological nature
 and are exactly calculable in many cases. Thus, centrally extended 
 supersymmetric theories present the first example of four-dimensional QFT 
 in which the physical masses (tensions) are known exactly, even at strong coupling.
 A great example is the mass formula for monopoles/dyons in the Seiberg-Witten solution \cite{sei}.

 \section{1990s and later}
 
 \subsection{Anomalous contribution to central charges}
 
 A mystery of the kink mass in the simplest two-dimensional ${\cal N}= (1,1)$ model
 \begin{eqnarray}
{\cal L} = \frac{1}{2}\left\{ \partial_\mu\phi \,\partial^\mu\phi
+\bar\psi\, i\! \!\not\!\partial \psi  -\left(\frac{\partial {\cal W}}{\partial\phi}
\right)^2
-\frac{\partial^2{\cal W}}{\partial\phi^2}\, \bar\psi\psi\right\}\, ,
\label{minlagp}
\end{eqnarray}
tortured theoretical physicists since 1983 for fifteen years. 
Here $\phi (t,z)  $ is a real scalar field, $\psi$ is a Majorana spinor,
 and ${\cal W}(\phi)$ is a superpotential.
 Superalgebra of this model is
 \begin{equation}
\{{Q}_\alpha, \bar Q_\beta\}
=2\,(\!\gamma^\mu)_{\alpha\beta} \,P_\mu +
2i\,(\!\gamma^5)_{\alpha\beta}\,
 {\cal Z}\;.
\label{ce}
\end{equation}
where $ {\cal Z}$ is the Lorentz-invariant central charge,
\begin{equation}
{\cal Z} =  \Delta {\cal W} \equiv \left( {\cal W}\right)_{z=+\infty} -\left( {\cal W}\right)_{z=-\infty}
\, .
\label{anocc}
\end{equation}
 
 In this model kinks belong to a short BPS-protected multiplet and, therefore,
 their mass was expected to be exactly equal to  ${\cal Z} $.
 However, in numerous calculations (for a review see \cite{Shifman}) this equality was found to be  violated already at one loop!  The solution was found only 
in 1998 \cite{Shifman}. It turned out that the equality $M_{\rm kink} =  \Delta {\cal W}$  could not be
valid because Eq. (\ref{anocc}) was incomplete. Nobody suspected the existence of a quantum anomaly
which replaces $\Delta{\cal W}$ in (\ref{anocc}) by
\beq
\Delta\left( {\cal W}+ \frac{1}{4\pi}\, {\cal
W}''\right)_{z=\pm\infty}\,.
\label{anoma1}
\eeq
After this replacement in (\ref{anocc}) the amended  expression for ${\cal Z}$ exactly coincides with the mass
$M_{\rm kink}$ in the short multiplet. 
 
 By the way, the model (\ref{minlagp}) has another anomaly -- global \cite{LSV}.
 The fermion field in (\ref{minlagp}) is real. Therefore, the fermion number $F$ is not defined, of course.
 However, looking at (\ref{minlagp}) one would say that the fermion parity $(-1)^F$ is well defined and conserved.
 This is a false impression. While conservation of $(-1)^F$ is valid in perturbation theory,
 it is lost nonperturbatively. Say, for a cubic superpotential (leading to a double-well potential) $(-1)^F$ ceases to be well-defined 
 for the kink supermultiplet because of a global anomaly discovered in \cite{LSV}. The short kink supermultiplet consists 
 of a single state for which the fermion parity is neither 1 nor $-1$.
 
 \subsection{The brane charge in \boldmath{${\cal N}=1$} pure Yang-Mills is anomaly }
 
 In Sect. \ref{22} I outlined the exact calculation of the gluino condenate carried out in the 1980s,
 \beq
2  {\rm Tr} \,\left\langle \lambda^\alpha \lambda_\alpha\right\rangle =
\langle
\lambda^{a}_{\alpha}\lambda^{a\,,\alpha}
\rangle = -6 N\Lambda^3 \exp \left({\frac{2\pi i k}{N}}\right)\,,
\,\,\, k = 0,1,..., N-1\,.
\label{wall15}
 \eeq
 The above result refers to ${\cal N}=1$ super-Yang-Mills.
 In 1996 Dvali and I found \cite{DvSh} a fascinating application for this result. 
  The $(1,0)$ and  $(0,1)$ branes charges below 
  \begin{equation}
\left\{Q_\alpha\,,Q_\beta\right\}=-4\,\Sigma_{\alpha\beta}\,\bar {
Z}\,,
\label{cccdw1}
\end{equation}
where 
\begin{equation}
\Sigma_{\alpha\beta}=-\frac 12\int {\rm d} x_{[\mu} {\rm d} x_{\nu ]}\,
(\sigma^\mu)_{\alpha\dot\alpha} (\bar \sigma^\nu)_{\beta}^{\dot\alpha}\,,
\end{equation}
are not seen at the classical level in this theory.  Nevertheless, they
exist  \cite{DvSh} as a quantum 
anomaly. They are saturated by domain walls interpolating
between vacua with distinct values of the  parameter $k$ in Eq. (\ref{wall15}),
labeling $N$ distinct vacua of 
super-Yang--Mills theory
with the gauge group SU($N$). The tension of the BPS wall is
\begin{equation}
T =|Z|= \frac{N}{8\pi^2} \left| \langle {\rm Tr} \lambda^2\rangle_{\rm vac \,\,f}
-\langle {\rm Tr} \lambda^2\rangle_{\rm vac \,\,i}
\right|
\label{eqte}
\end{equation}
where vac$_{\rm i,f}$ stands for the initial (final) vacuum between which the given
wall interpolates. This anomaly is in fact a ``superpartner" to that in the trace of the energy-momentum tensor.

For the interpolations between the neighboring vacua $\Delta{\rm Tr} \langle\lambda^2\rangle $ scales as $N^0$. Equation (\ref{eqte}) implies then that
the wall tension scales as $N^1$. Since the string coupling constant $g_s\sim 1/N$, the wall tension is proportional to $T\sim 1/g_s$ rather than $1/g_s^2$. Thus, this is not a ``normal"
soliton but, rather, a  D brane. 
This is the essence of
Witten's argument why the above walls should be considered
as analogs of D branes \cite{Witten}.

Many interesting consequences ensued. One of them was the Acharya--Vafa\index{Acharya--Vafa}\index{Vafa}
derivation of the wall world volume theory \cite{Achar}. Using a wrapped $D$-brane
picture and certain dualities they identified the $k$-wall\,\footnote{Minimal, or elementary,  walls\index{elementary walls} interpolate between 
vacua $n$ and $n+1$, while $k$-walls\index{k-walls} interpolate
between $n$ and $n+k$.} world volume theory
as 1+2 dimensional U($k$) gauge theory with the field content of
${\mathcal N}=2$ and the Chern--Simons term 
at level $N$ breaking ${\mathcal N}=2$ down to ${\mathcal N}=1$.
This allowed them to calculate the  wall multiplicity.

In fact,  even if  we consider a given minimal wall, $k_f=k_i+1$, we deal with several walls, all having one and the same tension.
The fact that distinct BPS walls with the same boundary conditions can have one and the same tension
is specific for supersymmetry. It was discovered (see \cite{kovner} and \cite{svritz,svritz2}) in studies of the BPS-saturated walls  --
in such walls, even if their internal structures are different,
the tension degeneracy is the consequence of the general law $T=|Z|$.

In the field-theoretic setting 
the $k$-wall
multiplicity was derived in \cite{svritz},
\begin{equation}
\nu_k = C_N^k=\frac{N!}{k! (N-k)!}\,.
\label{wmult}
\end{equation}
In particular, for the neighboring walls $k=1$ and hence $\nu = N$.
  The derivation is based on the fact
that the index\index{index} $\nu$ is topologically stable -- continuous deformations of the theory
do not change $\nu$. Thus, one can add an appropriate set of
matter fields sufficient for complete Higgsing of super-Yang-Mills. Upon Higgsing one can calculate the multiplicity $\nu_k $
at weak coupling.
It is worth noting that the method suggested in  \cite{svritz,svritz2} was recently extended \cite{ritz} to softly broken SQCD  at  $\theta =\pi$
with the purpose of exploring a newly discovered discrete anomaly 
\cite{gai}. 

The anomaly similar to (\ref{eqte}) was also found in two-dimensional $CP(N-1)$ models \cite{Losev} (see also \cite{book}). In this case it is a {\em bona fide} central charge, since the domain wall reduced to two dimensions becomes a kink (i.e. a localized particle).
Since the two-dimensional $CP(N-1)$ models with various degree of supersymmetry occur on the world sheet of non-Abelian strings (see below) this anomaly was extensively used in explorations of the non-Abelian strings \cite{used}. I will say more on this in the subsequent sections.

\subsection{Planar equivalence between \boldmath{${\cal N}=1$} super-Yang-Mills and its non-supersymmetric daughters}

In connection with the success of the exact predictions in supersymmetric theories a question was raised as to whether
one can draw quantitative lessons for nonsupersymmetric theories. The genesis of this question and first advances are described in detail in the review paper \cite{armoni}. In 2003 we noted and proved that ${\cal N}=1$ super-Yang-Mills
has two orientifold daughters which are perfectly similar to the parent supersymmetric theory except they are {\em not} supersymmetric!
Namely, the gluino field (described by a Weyl fermion in the adjoint representation) can be replaced by a Dirac fermion
in the two-index representation of $SU(N)$ -- either symmetric or antisymmetric. In the large-$N$ limit all correlation functions in the common sector in these two theories are equal \cite{ASV}. In practice, of most interest is the antisymmetric two-index representation because in the $N=3$  case   ${\cal N}=1$ super-Yang-Mills' daughter is just ``our" conventional QCD with one flavor. 

Using this fact and extrapolating from $N=\infty$ to $N=3$ we were able to obtain the first analytic prediction for the
quark condensate \cite{ASV2} which agrees well with experimental data and lattice simulations. 

The ``new'' large-$N$ limit (the ``ASV'' limit) in QCD with the two-index antisymmetric fermions
(alternative to that of 't Hooft) was  rather extensively applied to QCD phenomenology, see e.g.
\cite{Cohen}. I want to mention that in the ASV limit, unlike the 't Hooft limit, the widths of exotic mesons, such as tetraquarks,
die off at large $N$ \cite{Cohen2}.

%\newpage

\section{Non-Abelian strings}
\label{nas}

Since 2003 I work with Alexei Yung on making relatively realistic theoretical constructions addressing QCD-type confinement based
on supersymmetric results and inspired by the Seiberg-Witten solution \cite{sei}. As Seiberg and Witten, Alexei and I started from consideration of  ${\cal N}=2$ Yang-Mills theories in four dimensions, which may or may not  be deformed by breaking parameters
down to ${\cal N}=1$. Unlike \cite{sei} our main focus was on the so-called quark vacua. 

In this section I will present a general idea of what is now known as non-Abelian strings and some key results such as the exact 2D-4D correspondence.
This theme is still very much ``work in progress.'' Therefore, a more complete summary will appear later. 

Non-Abelian string is a vortex type soliton supported in four-dimensional Yang-Mills theories. In a class of such theories flux tube solutions were found in the 1980s, but they are {\em not} non-Abelian. So is the string in the Seiberg-Witten solution at small deformations $\mu$ (and only at small $\mu$ the solution is possible). All the above vortex strings are similar to the
Abrikosov-Nilsen-Olesen string \cite{ANO}. They are formed at low energies by ``photons" after electric charge condensation. 
All other gauge bosons acquire large masses and do not play a direct role in the string formation.

\begin{figure}[h]
\centerline{\includegraphics[width=10.25cm]{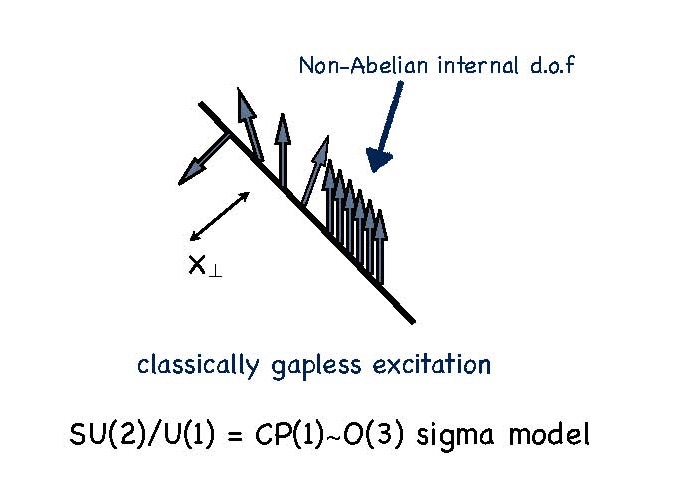}}
\caption{\small Orientational moduli on the string world sheet, see Sect. \ref{42}.}
\label{d11} 
\end{figure}

\begin{figure}[h]
\centerline{\includegraphics[width=10.25cm]{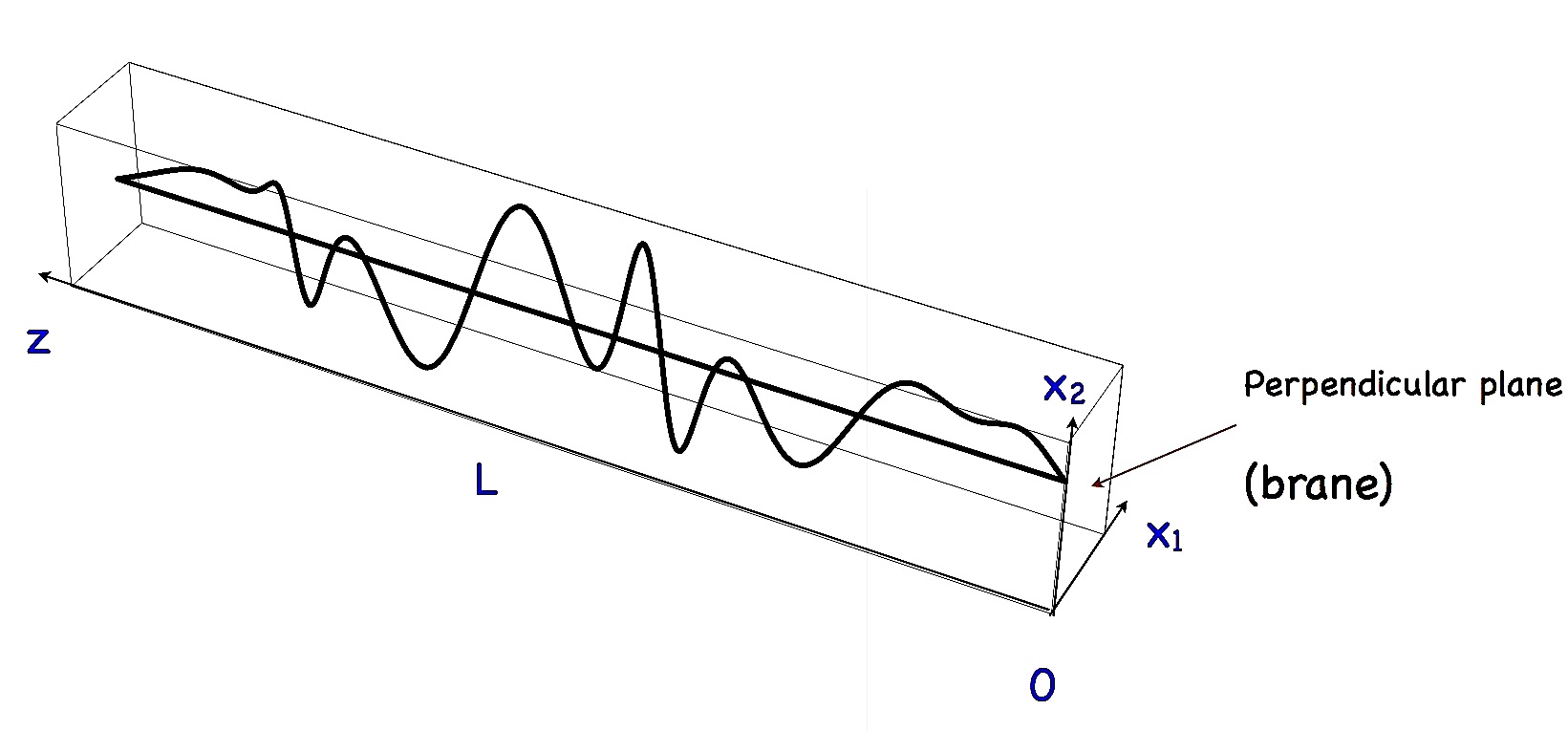}}
\caption{\small Excitations of translational moduli described by the Nambu-Goto action.}
\label{tran} 
\end{figure}

By non-Abelian string we mean a vortex soliton the structure of which is determined  on equal footing
by {\em all} gauge bosons existing in the bulk
(i.e. four-dimensional) theory. Correspondingly, there are no singled out $U(1)$ directions. As a result, new type of moduli on the string world sheet appear, reflecting arbitrariness of ``color"  orientations in the bulk gauge group,\footnote{Two-dimensional quantum oscillations
may and usually do provide a mass gap to these moduli which we refer to as ``orientational."}  see Fig. \ref{d11}. The orientational moduli, combined with the conventional translational moduli (Fig. \ref{tran}), form a dynamically nontrivial sigma model on the string world sheet.
As we will see shortly, studying dynamics in two dimensions one can project two-dimensional results for protected quantities onto four dimensions. 

\subsection{Basic Model}
\label{basic}

Chronologically first (and still most convenient) are the bulk ${\cal N}=2$ Yang-Mills theories with the gauge group $U(N)$ and $N_f=N$ ($N_f$ is the number of flavors), and the 
Fayet-Iliopoulos term. The Higgs branch is lifted by the Fayet-Iliopoulos parameter $\xi\neq 0$, and the quark vacua are isolated \cite{ayung}.

 The simplest and most pedagogical example is $N=N_f=2$.
  The bosonic part of the
theory has the form 
\beqn
S&=&\int d^4x \left[\frac1{4g^2_2}
\left(F^{a}_{\mu\nu}\right)^2 +
\frac1{4g^2_1}\left(F_{\mu\nu}\right)^2
+
\frac1{g^2_2}\left|D_{\mu}a^a\right|^2 +\frac1{g^2_1}
\left|\partial_{\mu}a\right|^2 \right.
\nonumber\\[4mm]
&+&\left. \left|\nabla_{\mu}
q^{A}\right|^2 + \left|\nabla_{\mu} \bar{\tilde{q}}^{A}\right|^2
+V(q^A,\tilde{q}_A,a^a,a)\right]\,.
\label{qed}
\eeqn
Here $\nabla_{\mu}$ is the covariant derivative in the fundamental representation
of  $SU(2)$, and
\beq
\nabla_\mu=\partial_\mu -\frac{i}{2}\; A_{\mu}
-i A^{a}_{\mu}\, T^a\,, \qquad T^a= \frac 12 \tau^a\,,
\label{defnabla}
\eeq
while $q^A$ and $\tilde{q}_A$ are quark hypermultiplets ($A=1,2$). The covariant derivative in the adjoint representation is denoted by $D_\mu$.
The coupling constants $g_1$ and $g_2$
correspond to the $U(1)$  and  $SU(2)$  sectors, respectively.
With our conventions, the $U(1)$ charges of the fundamental matter fields
are $\pm1/2$. The conventions in (\ref{qed}) are Euclidean. Moreover, $q$ and $\tilde q$ 
are the lowest (squark) components of the chiral superfields $Q$ and $\tilde Q$. Each flavor is composed of one $Q$ and one $\tilde Q$.
The scalar complex field $a$ is an ${\cal N}=2$ superpartner of the gauge fields.

\vspace{1mm}

The potential $V(q^A,\tilde{q}_A,a^a,a)$ in the action (\ref{qed})
is 
\beqn
&& V(q^A,\tilde{q}_A,a^a,a)  = 
 \frac{g^2_2}{2}
\left( \frac{i}{g^2_2}\,  \varepsilon^{abc} \bar a^b a^c
 +
 \bar{q}_A\,T^a q^A -
\tilde{q}_A T^a\,\bar{\tilde{q}}^A\right)^2
\nonumber\\[3mm]
&&+ \frac{g^2_1}{8}
\left(\bar{q}_A q^A - \tilde{q}_A \bar{\tilde{q}}^A \right)^2
+ 2g^2_2\left| \tilde{q}_A T^a q^A \right|^2+
\frac{g^2_1}{2}\left| \tilde{q}_A q^A -\xi \right|^2
\nonumber\\[3mm]
&&+\frac12\sum_{A=1}^N \left\{ \left|(a +2T^a a^a)q^A
\right|^2
+
\left|(a +2T^a a^a)\bar{\tilde{q}}^A
\right|^2 \right\}\,.
\label{pot}
\eeqn
Here   
the sum over the repeated flavor indices $A$ is implied. In the simplest version the mass terms of the matter hypermultiplets are omitted.

Let us discuss the vacuum structure of this model. 
We will limit ourselves 
to  isolated vacua with the maximal possible value
of condensed quarks -- two. 
The  vacua of the theory (\ref{qed}) are determined by the zeros of 
the potential (\ref{pot}). It is easy to see that the adjoint fields do not develop vacuum expectation values (VEVs),
\beq
\langle \Phi\rangle = 0
\label{avev}
\eeq
where we defined the scalar adjoint matrix as
\beq
\Phi = \frac12\, a + T^a\, a^a.
\label{Phidef}
\eeq
On the contrary, 
the squark fields do develop  VEVs which have the color-flavor locked  form
(up to possible gauge transformations)
\beqn
\langle q^{kA}\rangle &=&\langle \bar{\tilde{q}}^{kA}\rangle =\sqrt{
\frac{\xi}{2}}\,
\left(
\begin{array}{cc}
1 &  0 \\
0 &  1\\
\end{array}
\right),
\qquad
k=1,2,\qquad A=1,2\, ,
\label{qvev}
\eeqn
where the squark fields is written as an $2\times 2$ matrix in 
the color and flavor indices. The choice (\ref{qvev}) makes the potential (\ref{pot}) vanish.
For the time being 
the Fayet-Iliopoulos parameter $\xi$ is assumed to be large, $\xi\gg\Lambda^2$.

The  vacuum field (\ref{qvev}) results in  the spontaneous
breaking of both gauge and flavor $SU($2$)$ symmetries.
A diagonal global $SU(2)$ survives, however,
\beq
{\rm U}(2)_{\rm gauge}\times {\rm SU}(2)_{\rm flavor}
\to {\rm SU}(2)_{C+F}\,.
\label{c+f}
\eeq
Thus, a color-flavor locking takes place in the vacuum.
The above condensates imply that the basic theory under consideration is
fully Higgsed and is at weak coupling provided  $\xi$ is large.

\subsection{Non-Abelian Strings}
\label{42}

Why does the model described above support a novel type of strings, non-Abelian?
The ANO string corresponds to a U(1) winding of the phase of the squark fields in the plane, perpendicular to the
string axis,
\beq
q^{kA}\,\, \stackrel{\longrightarrow}{_{r\to\infty}}\,\, \sqrt{
\frac{\xi}{2}}\,e^{i\alpha}
\left(
\begin{array}{cc}
1 &  0 \\
0 &  1\\
\end{array}
\right),
\eeq
where $\alpha$ is the polar angle in the perpendicular plane. Its topological stability is due to $\pi_1({\rm U}(1)) = Z$.
Now we have more options. It is well known that $\pi_1({SU}(2)) $ is trivial and, seemingly, there are no topologically stable strings other than those due to $\pi_1({U}(1)) $, i.e. the good old Abelian strings.

This is not the case, however. Observe that the center of the SU(2) group, $Z_2$, belongs to the U(1) factor too. This means
that we  need 
$\pi_1({SU}(2)\times U(1)/Z_2) $ nontrivial. It is easy to see that $\pi_1({SU}(2)\times U(1)/Z_2) =Z_2$. We can built a topologically stable $Z_2$ string.

One can split the $2\pi$ windings in two halves: the first (from 1 to $-1$) is carried out in $U(1)$, while the second,
from $-1$ to 1 in $SU(2)$ (e.g. around the third axis in the ``isospace").  Correspondingly, the winding ansatz takes the form
\beq
q^{kA}\longrightarrow\sqrt{
\frac{\xi}{2}}
\left(
\begin{array}{cc}
\,e^{i\alpha}  &  0 \\
0 &  1\\
\end{array}
\right)\quad {\rm or} \quad 
q^{kA}\longrightarrow\sqrt{
\frac{\xi}{2}}
\left(
\begin{array}{cc}
1 &  0 \\
0 &  \,e^{i\alpha}\\
\end{array}
\right),
\label{12p}
\eeq
depending on which of the two  combination of generators $T_{{\rm U}(1)} \pm T^3_{{\rm SU}(2)}$
we use.

It is clear, that the ansatz (\ref{12p}) breaks the color-flavor locked SU(2) down to U(1).
The particular way of embedding is unimportant. Instead of $T^3$ we could have chosen any other generator of $SU(2)$.
In other words, the existence of the string based on (\ref{12p}) implies the existence of the whole family of strings parametrized by
the coset $SU(2)/U(1)$ moduli. Then, the theory of the moduli fields on the string world sheet is obviously the $CP(1)$ model. It is asymptotically free in the ultroviolet (UV)  and strongly coupled in the infrared (IR). 
Since the string is 1/2 BPS saturated, the world-sheet model has four supercharges. 
Thus, we arrive at the ${\cal N}= (2,2)$ model. 
The existence of the orientational moduli implies that the flux through the string does not have a preferred orientation inside $SU(2)$. This string is genuinely non-Abelian.

If the tension of the ANO string is $4\pi\xi$, the tension of the non-Abelian string is $2\pi\xi$. 
For arbitrary $N$ the tension of the ANO string is $2N\pi\xi$, while the tension of our non-Abelian string remains $2\pi\xi$; it is $N$-independent.  This is an important consequence of the BPS saturation.

\subsection{Kinks as Confined Monopoles}

There are two degenerate vacua in ${\mathcal N}=2$ $CP(1)$ model. This is dictated by the Witten index. The 
existence of two isolated vacua are not seen classically, since classically we do not see mass gap generation. They are
labeled by the fermion condensate $\langle\bar\psi\psi\rangle$, which can take two values, in much the same way as the gluino condensate in ${\cal N}=1$ $SU(2)$ Yang-Mills theory.

\begin{figure}[h]
\centerline{\includegraphics[width=8cm]{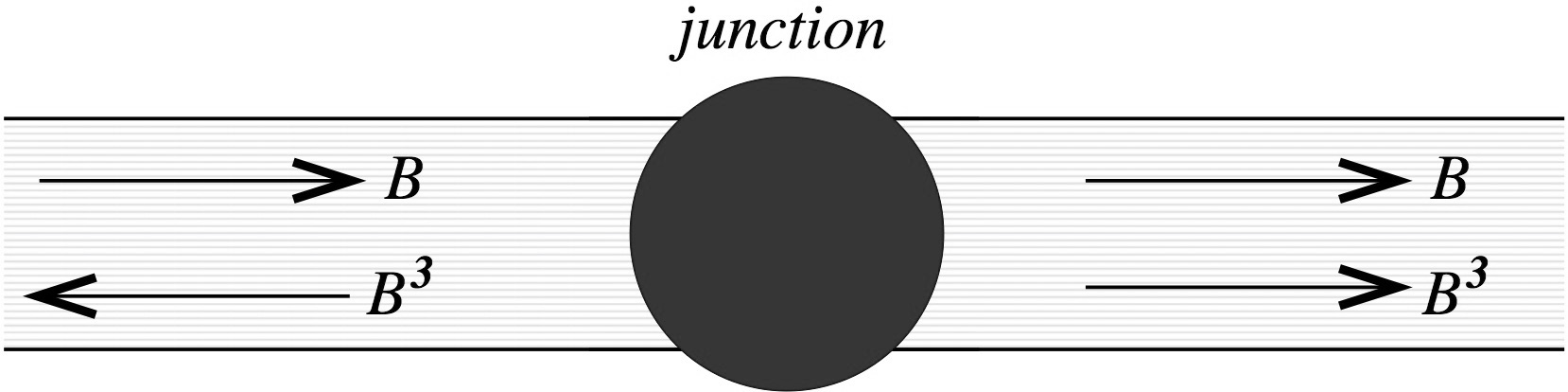}}
\caption{\small $CP(1)$ model kink as a junction of two degenerate strings.}
\label{sixf} 
\end{figure}

From the bulk standpoint this means that there exist two distinct strings, both with the tension value $2\pi\xi$. If so, there should exist a junction of these two strings. Returning to two dimensions we can say that this junction is a kink interpolating between two distinct vacua of the supersymmetric $CP(1)$ model
(Fig. \ref{sixf}). 

The kink itself is BPS saturated since, as I mentioned above, 
${\mathcal N}=(2,2)$ algebra in 
$CP(1)$ has a central extension. Therefore, the mass of the kink is exactly calculable. 

Moreover, using supersymmetry one can prove that this two-dimensional kink is a reincarnation of a confined monopole, i.e. a monopole with strings attached to it. To this end we must use exact holomorphic dependences protected by supersymmetry.

\begin{figure}[h]
\centerline{\includegraphics[width=12cm]{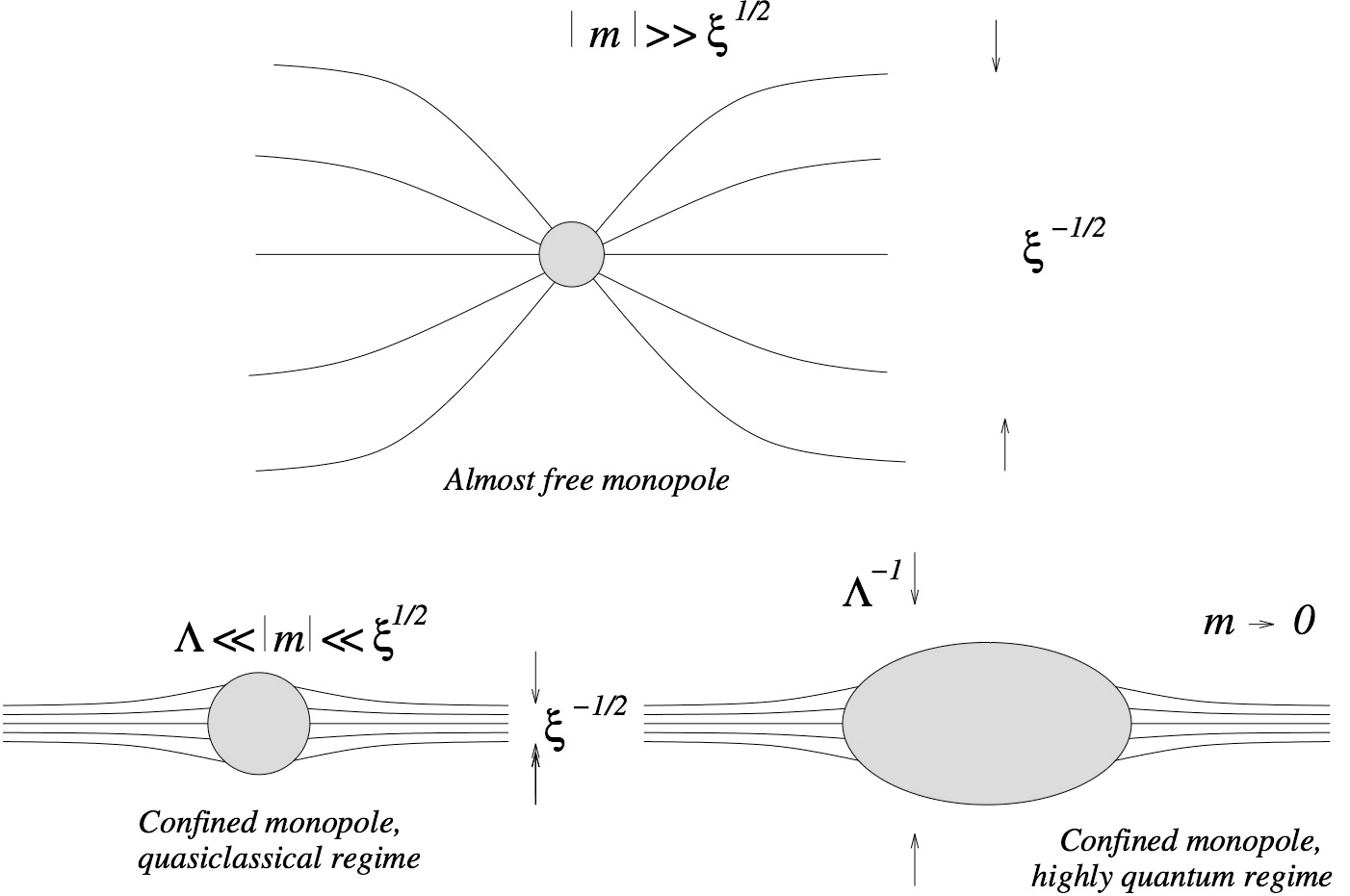}}
\caption{\small Evolution of the confined monopoles.}
\label{sixfp} 
\end{figure}

Since so far we set the quark mass term $m=0$, the theory under consideration has two parameters of dimension of mass: the string tension $\xi$ and the dynamical scale $\Lambda$. Note a ``miraculous" fact that $\Lambda$ is one and the same in the bulk theory and in the string world-sheet model! In the limit $m_1=m_2=0$ the kink is {\em unrecognizable} as a 4D monopole. However, if we introduce another holomorphic parameter of dimension of mass 
(the so-called twisted mass which preserves ${\cal N}=2$)
and allow it to vary continuously, we will see that starting 
from the classical 't Hooft--Polyakov monopole, and analytically continue (smoothly, with no singularities) this parameter, we arrive at the kink on the string. 

The crucial parameter is the squark mass difference $\Delta m_{12}= m_1-m_2\equiv 2m\neq 0$. It is clear that $\Delta m_{12} $ 
is a complex parameter, and so is $\Lambda$.  At the same time $\xi$ is a real parameter, and, as a result, the appropriate central charges cannot depend on $\xi$. At $\xi=0$ and $\Delta m_{12}\neq 0$ we are on the Coulomb branch.

 If $|m|\gg \xi$ (or, alternatively, $\xi\to 0$), we arrive (in the bosonic sector) at the Georgi-Glashow model, with the classical 't Hooft--Polyakov monopoles. Making $\xi$ nonvanishing but small we attach strings to the 't Hooft--Polyakov monopole, although these flux tubes form far away from its core, which (the core) remains easily recognizable. Evolving towards smaller values of $|m|$ and larger values of $\xi$ we arrive, at the end of the evolution process, at the kink described above. Thus, indeed, this kink is the monopole's apparent heir.
Figure \ref{sixfp} illustrates this analytic continuation.

Four-dimensional expression for the free monopole mass (i.e. on the Coulomb branch at $\xi =0$) obtained in \cite{sei}
in our problem reduces to 
\beq
M_{\rm mon}^{\rm Coulomb} =\sqrt{2}\left| a_D^3 \left( a^3 = - \frac{\Delta m_{12}}{\sqrt 2}\right)\right|,
\label{mon29}
\eeq
where $a_D^3$ is the dual Seiberg-Witten potential for the $SU(2)$ gauge group. Now, if one switches on a Fayet-Iliopoulos  parameter $\xi\neq 0$, {\em a priori }
one could expect  corrections to
the monopole mass  depending on $\sqrt{\xi}/\Lambda$
but  in fact they are {\em forbidden}
by U(1)$_R$ charges. Therefore,
\beq
M_{\rm mon}^{\rm Coulomb }= M_{\rm mon}^{\rm confinement }\,.
\label{mmCc}
\eeq
On the other hand, the monopole mass in the confinement phase
is given by the  kink mass  in the $CP(1)$ model, 
\beq
M_{\rm kink} =\left| Z_{\rm kink}\right| 
\label{mon30}
\eeq
where 
$Z_{\rm kink}$ is given by the following formula:
\beq
\big\langle Z \big\rangle_{{\rm kink}}= -\frac{i}{2\pi}
\left\{\Delta m_{12} \log\frac{\Delta m_{12}+\sqrt{\Delta m_{12}^2+4\Lambda^2} }
{\Delta m_{12}- \sqrt{\Delta m_{12}^2+4\Lambda^2}} -
2\sqrt{\Delta m_{12}^2+4\Lambda^2}\right\}.
\label{ndef}
\eeq
If the two-dimensional kink indeed represents a confined four-dimensional monopole,
then Eq. (\ref{mon30}) can be extended,
\beq
M_{\rm mon}^{\rm Coulomb }\leftrightarrow M_{\rm mon}^{\rm confinement }
\leftrightarrow M_{\rm kink}\,.
\label{mmk}
\eeq

Comparing (\ref{mon29}) and (\ref{ndef}) we confirm exact agreement and thus establish the 2D-4D correspondence.
This and other cases of the 2D-4D correspondence are reviewed in
\cite{book,rn1}.

A remarkably close parallel between
four-dimensional $SU(2)$ Yang--Mills theory with $N_f=2$ on the Coulomb branch and the two-dimensional
$CP(1)$ model was noted  in \cite{Dorey}. The coincidence observed by Dorey 
remained mysterious and could not be understood before the advent of non-Abelian strings.

\subsection{\boldmath${\cal N}=1$ bulk theories give rise to \boldmath${\cal N}=(0,2)$ sigma models on the world sheet}

If we deform the basic model described in Sec. \ref{basic} by adding $\mu\,  {\rm Tr}( a^2)$ term in the superpotential
(this term breaks the bulk symmetry down to ${\cal N}=1$) then a nonminimal {\em heterotic} $CP(N-1)$ model emerges on the string world sheet \cite{het1,het2}. This is a remarkably rich two-dimensional model.
Supersymmetry on the world sheet is spontaneously broken \cite{het3} in this model. At the same time there are still $N$ degenerate vacua and, therefore,
the kinks representing confined monopoles survive. The world-sheet dynamics becomes highly nontrivial if we add
twisted masses. In this case the large-$N$ solution exhibits three distinct phases \cite{het3} depending on the ratio $m/\Lambda$. 

\section{A non-Abelian string can be critical}

Most of the non-Abelian strings we constructed and explored are not ultraviolet-complete in the sense of Polchinski and Strominger \cite{PS}. This means that they cannot be quantized as strings at energies higher than the
inverse thickness of the vortex string at hand. It is clear that at such energies we have 
to take into account higher derivative corrections. 
 The blow up of higher derivative terms in
the world-sheet theory reflects a finite thickness of the solitonic vortex strings. Some time ago a question was raised \cite{cs1}
(see also \cite{cs2}) whether one can find a solitonic string which is critical. To this end it must be a ten-dimensional 
superstring, of course. 

If such solitonic string exists it must satisfy a number of conditions, namely, 

(i) it should be infinitely thin, implying that higher derivative terms on the world sheet can be ignored;

(ii) The world-sheet theory must be conformally invariant;

(iii) The theory must have the critical value of the Virasoro central charge.

The conditions  (i), (ii) and (iii) above are met by the non-Abelian vortex string 
\cite{ayung} supported in four-dimensional 
\ntwo supersymmetric QCD with the U$(N)$ gauge group, $N_f=2N$ matter hypermultiplets  and the
Fayet-Iliopoulos parameter $\xi$. To ensure the appropriate value of the Virasoro central charge and make the 
four-dimensional problem effectivly ten-dimensional we 
need   $N=2$ and $N_f=4$.

The non-Abelian vortex string meeting the above conditions has \ntwot super\-symmetry on its world sheet.
In contradistinction to the $N=N_f$ case discussed in Sect. \ref{nas} 
it has not only translational and orientational moduli, but the so-called size moduli as well \cite{sl}.
This is due to the fact that the dynamics of orientational and size moduli 
are described by two-dimensional sigma model 
known in physics as  $WCP(N,N_f-N)$ model\,\footnote{Both the orientational and the size moduli
have logarithmically divergent norms, see e.g.  \cite{sl}. After an appropriate infrared 
regularization, logarithmically divergent norms can be absorbed into the definition of 
relevant two-dimensional fields.
In fact, the world-sheet theory on the semilocal non-Abelian string is 
not exactly the $WCP(N,N_F-N)$ model \cite{SVY}, there are minor differences inessential in the infrared.}  and in mathematics as
\beq
\mathcal{O}(-1)^{\oplus(N_f-N)}_{\mathbb{CP}^1}\,.
\label{12}
\eeq
 For $N_f=2N$
the model is conformal and the condition (ii) above is satisfied. Moreover for $N=2$ the 
dimension of orientational/size moduli space is six and they can be combined with 
four translational moduli to form a ten-dimensional space required for critical superstrings.\footnote{It corresponds to $\widehat{c}=\frac{c}{3}=3$.}
Thus the condition (iii) is  satisfied too \cite{cs1}.
For $N=2$ the  sigma model target space is a six-dimensional
non-compact Calabi-Yau manifold $Y_6$, namely, the resolved conifold.

Given that the conditions (ii) and (iii) are met, we assumed \cite{cs1} that
vortex string satisfies the thin-string condition (i) at strong coupling,
\beq
\ell \to 0\,, \,\,\, \mbox{at}\,\,\, g^2\to g_c^2 \sim 1\,,
\eeq
where $\ell$ is the string thickness. 

As well known \cite{ArgPlessShapiro,APS}, the bulk theory at hand possesses a strong-weak coupling 
duality.\footnote{Argyres et al. proved this duality for $\xi =0$. It should allow one to 
study the bulk theory at strong coupling in terms of weakly coupled dual theory  
 at $\xi\neq 0$ too.}
So does the two-dimensional world sheet theory: the $WCP(N,N_f-N)$ model  is  
self-dual under the reflection of the coupling constant $\beta$,
\beq
\beta \to  \beta_D = -\beta\,.
\label{CPduality}
\eeq
Under transformation (\ref{CPduality})  the orientational and size moduli  interchange.
Note that the 2D coupling constant $\beta$ can be complexified by including the $\theta$ term in the 
action of the ${CP}(N-1)$ model, 
$$\beta \to \beta + i\,\frac{\theta_{2D}}{2\pi}\,.$$ 
As a result, one can derive an exact 2D-4D map between the coupling constants, see  \cite{Kom} for a detailed analysis. 
The 4D self-dual point $g^2=4\pi$ is mapped onto the 2D self-dual point $\beta=0$.

The thin string hypothesis is equivalent
to the following statement
\beq
\ell^{\,-2}  \to \xi\times 
\left\{
\begin{array}{ccc}
g^2, & g^2\ll 1& \\
\infty, & g^2\to 4\pi& \\
16\pi^2/g^2,& g^2\gg 1 & \\
\end{array},
\right.
\label{msing}
\eeq
where the dependence of $\ell^{-2}$  at small and large $g^2$ follows from the weak coupling formula 
for the Higgsed bulk gauge bosons and
duality \cite{ArgPlessShapiro,APS}. In terms of $\beta$ the critical point is $\beta=0$. At this point the target space of the 
$WCP(2,2)$ part of the world-sheet model develops a conical singularity. The number of real (bosonic) degrees of freedom parametrizing $WCP(2,2)$  is six. Adding four translational moduli we get ten-dimensional space
The critical string we arrived at can be viewed as a type IIA superstring,
 a version of the Kutasov-Vafa little string. The target space is $\mathbb{R}^4\times WCP(2,2)= \mathbb{R}^4\times Y_6$ where $Y_6$ is a non-compact  
Calabi-Yau conifold. 

\section{Briefly about spectrum}

\subsection{Massless states}
\label{mst}

First, we addressed  massless four-dimensional excitations of the quantized string. To this end zero modes of appropriate operators
in the $Y_6$ background were found. At first sight one might think that there are no {\em normalizable} zero modes at all, because our
our Calabi-Yau space is {\em non-compact}.
As a matter of fact, at the selfdual value of $\beta=0$ a marginally normalizable scalar zero mode exists!

Our analysis led us to the conclusion
that the only road leading to the above zero mode is as follows:
\beq
\delta G_{ij}=\phi_4(x)\,\delta g_{ij}(y)\,,
\label{factor}
\eeq
where $x_{\mu}$ and $y_i$ are the coordinates on $\mathbb{R}^4$ and $Y_6$, respectively, and $G_{ij}$ is the metric on 
$Y_6$. Then we studied the relevant Lichnerowicz equation on $Y_6$ \cite{cs1,cs2}. 
Solutions of this equation for the Calabi-Yau spaces reduce to  deformations of the K\"ahler form or deformations
of complex structure \cite{NVafa,GukVafaWitt}. In the former case we deal with the {\em resolved} conifold while in the latter case with the {\em deformed} conifold. The difference between the two lies in the method of smoothing
the conifold singularity. For the resolved conifold one introduces 
a non-zero (but small) $\beta$  preserving 
the K\"ahler structure and Ricci-flatness of the metric. 
The explicit metric for the resolved conifold can be found in 
\cite{Candel,Zayas,Klebanov}. The resolved conifold has no normalizable zero modes. In particular, 
it is demonstrated in \cite{cs2} that the four-dimensional scalar $\beta$ associated with the K\"ahler form
deformation is non-normalizable.

At the same time deformation of the complex structure \cite{NVafa} does lead to a marginally normalizable 
four-dimensional scalar localized on the string in the same sense as the orientational 
and size zero modes 
are localized on the vortex-string  solution. 

Thus, a four-dimensional  massless scalar $b$ is a part of a four-dimensional ${\cal N}=2$ hypermultiplet. This implies that 
we observe a new Higgs branch in the bulk which is developed only at the self-dual value of the bulk
coupling constant $g^2=4\pi$.

\subsection{Massive states}

The critical string theory on the conifold is hard
to use for calculating the spectrum of massive string modes because (unlike Sect. (\ref{mst})) the supergravity approximation
does not work. However, in the given problem one can act in a different way  and use the equivalent formulation of the
 theory as a non-critical   $c=1$ string theory with the Liouville field and a compact scalar at 
the self-dual radius \cite{GivKut,GVafa}. Non-critical $c=1$ string theory is formulated on the target space
\beq
\mathbb{R}^4\times \mathbb{R}_{\phi}\times S^1,
\label{target}
\eeq
where $\mathbb{R}_{\phi}$ is a real line associated with Liouville field $\phi$.

\begin{figure}
\epsfxsize=7cm
\centerline{\epsfbox{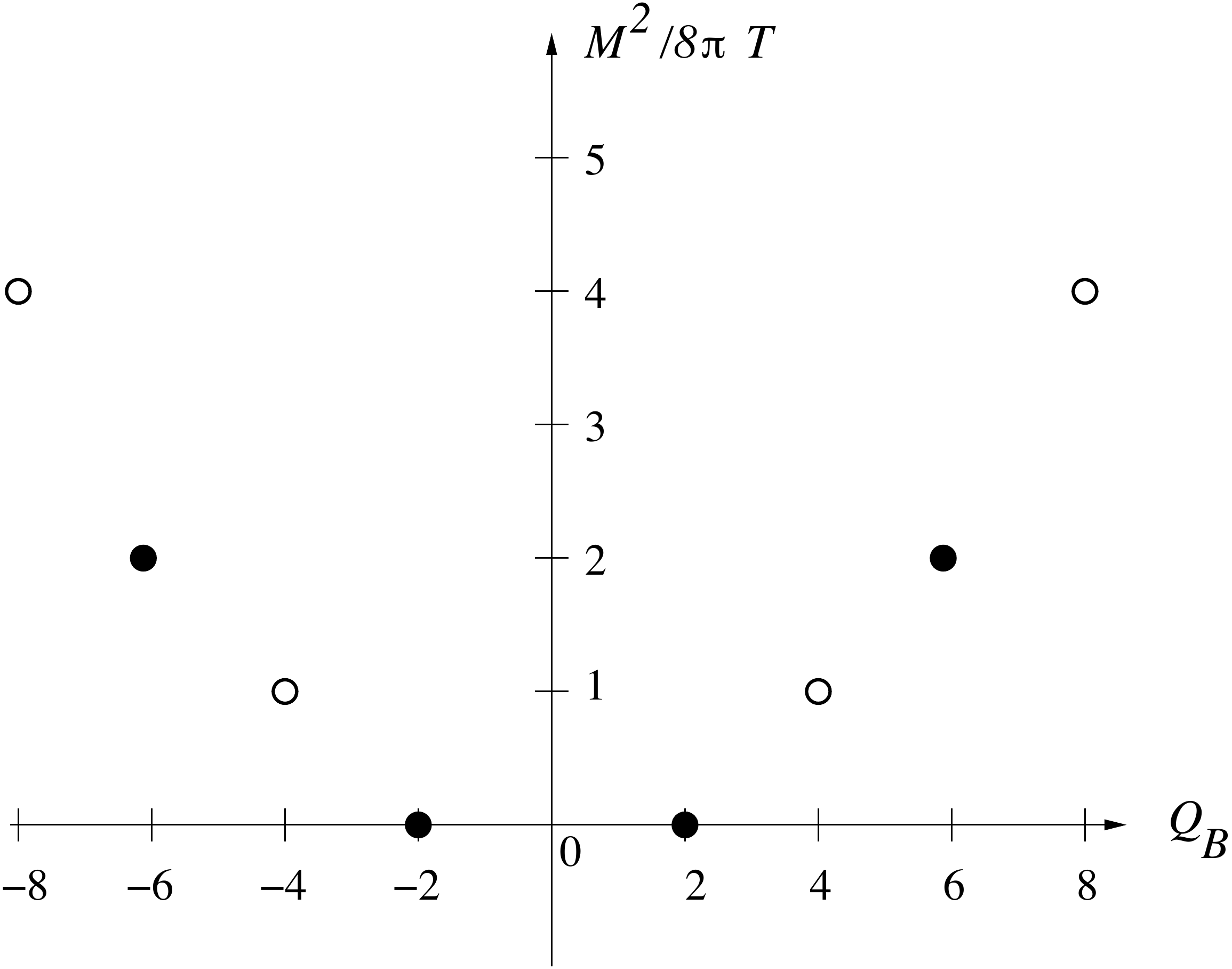}}
\caption{\small  Spectrum of spin-0 and spin-2 states as a function of the baryonic charge. Closed 
and open circles denote  spin-0 and spin-2 states, respectively.
 }
\label{fig_spectrum}
\end{figure}

Our first results on the spectrum of the solitonic-critical string \cite{sylbo} are presented in Fig.~\ref{fig_spectrum}
which shows the masses of 4D spin-0 and spin-2 states as a function of the 
baryonic charge.
The results of the current studies, including the 4D hypermultiplet structure, will be reported elsewhere \cite{tbp}.

\section{Instead of conclusions. A digression}

I will not summarize my talk for many reasons. My presentation is rather concise (although covers many topics on which I worked since 1980)
and hopefully accessible to the interested reader,
being supplemented by a detailed list of references below. Instead, I'd like to discuss a rather pessimistic question that was posed by
 Richard Feynman\cite{Fey} fifty years ago. Feynman wrote:

\begin{quote}

What will happen [to our science] ultimately? We are going along guessing the laws; how many laws are we going to have to guess? I do not know. Some of my colleagues say that this fundamental aspect of our science will go on; but I think there will certainly not be perpetual novelty, say for a thousand years. This thing cannot keep on going so that we are always going to discover more and more new laws. If we do, it will become boring that there are so many levels one underneath the other. It seems to me that what can happen in the future is either that all the laws become known -- that is, if you had enough laws you could compute consequences and they would always agree with experiment, which would be the end of the line -- or it may happen that the experiments get harder and harder to make, more and more expensive, so you get 99.9 per cent of the phenomena, but there is always some phenomenon which has just been discovered, which is very hard to measure, and which disagrees; and as soon as you have the explanation of that one there is always another one, and it gets slower and slower and more and more uninteresting. That is another way it may end. But I think it has to end in one way or another.

\end{quote}

Any thoughts?

\section*{Acknowledgments}

I am grateful to Alexei Yung for endless discussions. Correspondence with E.~Gerchkovitz, A.~Karasik and A. Kataev is acknowledged.
This work is supported in part by DOE grant de-sc0011842.


\vspace{1cm}

\begin{thebibliography}{99}

{\small
\bibitem{GL}
{\em Extension of the Algebra of Poincare Group Generators and Violation of P Invariance,}
  JETP Lett.\  {\bf 13}, 323 (1971), [reprinted in {\em Supersymmetry}, 
Ed.  S. Ferrara, (North-Hollands/World Sci, 1987), vol. 1, page 7].
  %%CITATION = JTPLA,13,323;%%
  %1274 citations counted in INSPIRE as of 21 Feb 2017

  \bibitem{PauliLec}
W. Pauli, {\sl Pauli Lectures on Physics}, Vol. 6, {\em Selected Topics on Field Quantization},
(MIT Press, Cambridge, MA, 1973), p. 33.

\bibitem{zve}
E. Likhtman, FIAN Report No. 41 (1971), [Reprinted in English in E. Likhtman, {\em About SuSy 1970}, in 
Proceedings of the Intern. Symp. ``Celebrating  30 Years of Supersymmetry," Eds. K. Olive, S. Rudaz and M. Shifman
(Nucl. Phys. B (Proc. Suppl.) 101 (2001)), pp. 5-14].

\bibitem{NSVZinst}
V.~A.~Novikov, M.~A.~Shifman, A.~I.~Vainshtein and V.~I.~Zakharov,
{\em Exact Gell-Mann-Low Function of Supersymmetric Yang-Mills Theories from Instanton Calculus,}
  Nucl.\ Phys.\ B {\bf 229}, 381 (1983).
  %doi:10.1016/0550-3213(83)90338-3
  %%CITATION = doi:10.1016/0550-3213(83)90338-3;%%
  %663 citations counted in INSPIRE as of 22 Feb 2017
  
    \bibitem{BPST} 
  A.~A.~Belavin, A.~M.~Polyakov, A.~S.~Schwartz and Y.~S.~Tyupkin,
{\em Pseudoparticle Solutions of the Yang-Mills Equations,}
  Phys.\ Lett.\  {\bf 59B}, 85 (1975).
%  doi:10.1016/0370-2693(75)90163-X
  %%CITATION = doi:10.1016/0370-2693(75)90163-X;%%
  %2601 citations counted in INSPIRE as of 31 Dec 2016
  
  \bibitem{NSVZc}
  V.~A.~Novikov, M.~A.~Shifman, A.~I.~Vainshtein and V.~I.~Zakharov,
 {\em Supersymmetric Instanton Calculus (Gauge Theories with Matter),}
  Nucl.\ Phys.\ B {\bf 260}, 157 (1985)
  %%CITATION = doi:10.1016/0550-3213(85)90316-5;%%
  %201 citations counted in INSPIRE as of 22 Feb 2017
  
 \bibitem{sak}
 M.~A.~Shifman,
{\em Exact results in gauge theories: Putting supersymmetry to work. The 1999 Sakurai Prize Lecture,}
  Int.\ J.\ Mod.\ Phys.\ A {\bf 14}, 5017 (1999)
 % doi:10.1142/S0217751X99002372
  [hep-th/9906049].
  %%CITATION = doi:10.1142/S0217751X99002372;%%
  %22 citations counted in INSPIRE as of 22 Feb 2017

\bibitem{sei}
  N.~Seiberg and E.~Witten,
  {\em Electric-magnetic duality, monopole condensation, and confinement in ${\cal N}=2$ supersymmetric Yang-Mills theory,}
  Nucl.\ Phys.\ B {\bf 426}, 19 (1994)
  Erratum: [Nucl.\ Phys.\ B {\bf 430}, 485 (1994)]
  %doi:10.1016/0550-3213(94)90124-4, 10.1016/0550-3213(94)00449-8
  [hep-th/9407087];
  %%CITATION = doi:10.1016/0550-3213(94)90124-4, 10.1016/0550-3213(94)00449-8;%%
  %2854 citations counted in INSPIRE as of 27 Jan 2017
  {\em Monopoles, duality and chiral symmetry breaking in  ${\cal N}=2$ supersymmetric QCD,}
  Nucl.\ Phys.\ B {\bf 431}, 484 (1994)
  %doi:10.1016/0550-3213(94)90214-3
  [hep-th/9408099].
  
  \bibitem{nsvzbetamatter}
   V.~A.~Novikov, M.~A.~Shifman, A.~I.~Vainshtein and V.~I.~Zakharov,
 {\em Beta Function in Supersymmetric Gauge Theories: Instantons Versus Traditional Approach,}
  Phys.\ Lett.\  {\bf 166B}, 329 (1986).
%  [Sov.\ J.\ Nucl.\ Phys.\  {\bf 43}, 294 (1986)]
%  [Yad.\ Fiz.\  {\bf 43}, 459 (1986)].
%  doi:10.1016/0370-2693(86)90810-5
  %%CITATION = doi:10.1016/0370-2693(86)90810-5;%%
  %266 citations counted in INSPIRE as of 22 Feb 2017
  
  \bibitem{ShifmanVainshtein}
M.~A.~Shifman and A.~I.~Vainshtein,
{\em Solution of the Anomaly Puzzle in SUSY Gauge Theories and the Wilson Operator Expansion,}
  Nucl.\ Phys.\ B {\bf 277}, 456 (1986).
%  [Sov.\ Phys.\ JETP {\bf 64}, 428 (1986)]
%  [Zh.\ Eksp.\ Teor.\ Fiz.\  {\bf 91}, 723 (1986)].
%  doi:10.1016/0550-3213(86)90451-7
  %%CITATION = doi:10.1016/0550-3213(86)90451-7;%%
  %505 citations counted in INSPIRE as of 22 Feb 2017
  
   \bibitem{Step}
   V.~Y.~Shakhmanov and K.~V.~Stepanyantz,
 {\em Three-loop NSVZ relation for terms quartic in the Yukawa couplings with the higher covariant derivative regularization,}
  Nucl.\ Phys.\ B {\bf 920}, 345 (2017)
 % doi:10.1016/j.nuclphysb.2017.04.017
  [arXiv:1703.10569 [hep-th]];
  %%CITATION = doi:10.1016/j.nuclphysb.2017.04.017;%%
  %6 citations counted in INSPIRE as of 02 Apr 2018
  A.~E.~Kazantsev, V.~Y.~Shakhmanov and K.~V.~Stepanyantz,
{\em New form of the exact NSVZ $\beta$-function: the three-loop verification for terms containing Yukawa couplings,}
  arXiv:1803.06612 [hep-th].
  %%CITATION = ARXIV:1803.06612;%%
   
    \bibitem{Ko}
  Z.~Komargodski and N.~Seiberg,
 {\em Comments on the Fayet-Iliopoulos Term in Field Theory and Supergravity,}
  JHEP {\bf 0906}, 007 (2009)
  %doi:10.1088/1126-6708/2009/06/007
  [arXiv:0904.1159 [hep-th]];
  %%CITATION = doi:10.1088/1126-6708/2009/06/007;%%
  %104 citations counted in INSPIRE as of 22 Feb 2017
{\em Comments on Supercurrent Multiplets, Supersymmetric Field Theories and Supergravity,}
  JHEP {\bf 1007}, 017 (2010);
%doi:10.1007/JHEP07(2010)017
  [arXiv:1002.2228 [hep-th]].
  %%CITATION = doi:10.1007/JHEP07(2010)017;%%
  %136 citations counted in INSPIRE as of 22 Feb 2017
T.~T.~Dumitrescu and N.~Seiberg,
{\em Supercurrents and Brane Currents in Diverse Dimensions,}
  JHEP {\bf 1107}, 095 (2011)
 % doi:10.1007/JHEP07(2011)095
  [arXiv:1106.0031 [hep-th]].
  %%CITATION = doi:10.1007/JHEP07(2011)095;%%
  %70 citations counted in INSPIRE as of 22 Feb 2017
  
  \bibitem{Seiberg}
  N.~Seiberg,
{\em Electric -- magnetic duality in supersymmetric non-Abelian gauge theories,}
  Nucl.\ Phys.\ B {\bf 435}, 129 (1995);
  %doi:10.1016/0550-3213(94)00023-8
  [hep-th/9411149].
  %%CITATION = doi:10.1016/0550-3213(94)00023-8;%%
  %1388 citations counted in INSPIRE as of 22 Feb 2017
  see also  K.~A.~Intriligator and N.~Seiberg,
{\em Lectures on supersymmetric gauge theories and electric-magnetic duality,}
  Nucl.\ Phys.\ Proc.\ Suppl.\  {\bf 45BC}, 1 (1996)
%  [Subnucl.\ Ser.\  {\bf 34}, 237 (1997)]
%  doi:10.1016/0920-5632(95)00626-5
  [hep-th/9509066].
  %%CITATION = doi:10.1016/0920-5632(95)00626-5;%%
  %684 citations counted in INSPIRE as of 22 Feb 2017
  
  \bibitem{sigma}
V.~A.~Novikov, M.~A.~Shifman, A.~I.~Vainshtein and V.~I.~Zakharov,
{\em Instantons And Exact Gell-Mann-Low Function Of Supersymmetric O(3) Sigma Model,}
  Phys.\ Lett.\  {\bf 139B}, 389 (1984);
 % doi:10.1016/0370-2693(84)91837-9
  %%CITATION = doi:10.1016/0370-2693(84)91837-9;%%
  %43 citations counted in INSPIRE as of 22 Feb 2017);
A.~Y.~Morozov, A.~M.~Perelomov and M.~A.~Shifman,
{\em Exact Gell-Mann-Low Function Of Supersymmetric K\"ahler Sigma Models,}
  Nucl.\ Phys.\ B {\bf 248}, 279 (1984).
  %doi:10.1016/0550-3213(84)90598-4
  %%CITATION = doi:10.1016/0550-3213(84)90598-4;%%
  %43 citations counted in INSPIRE as of 22 Feb 2017
  
  \bibitem{Moore}
    G.~W.~Moore and P.~C.~Nelson,
 {\em The Etiology of $\sigma$ Model Anomalies,}
  Commun.\ Math.\ Phys.\  {\bf 100}, 83 (1985).
  %doi:10.1007/BF01212688
  %%CITATION = doi:10.1007/BF01212688;%%
  %78 citations counted in INSPIRE as of 01 Apr 2018) we have \cite{cui1}
  
   \bibitem{Chen}
   J.~Chen, X.~Cui, M.~Shifman and A.~Vainshtein,
  {\em On isometry anomalies in minimal ${\cal N} = (0,1)$ and ${\cal N} = (0,2)$ sigma models,}
  Int.\ J.\ Mod.\ Phys.\ A {\bf 31}, no. 27, 1650147 (2016)
 % doi:10.1142/S0217751X16501475
  [arXiv:1510.04324 [hep-th]];
  %%CITATION = doi:10.1142/S0217751X16501475;%%
  %4 citations counted in INSPIRE as of 01 Apr 2018
  {\em Anomalies of minimal ${ \mathcal N }=(0,1)$ and ${ \mathcal N }=(0,2)$ sigma models on homogeneous spaces,}
  J.\ Phys.\ A {\bf 50}, no. 2, 025401 (2017)
  %doi:10.1088/1751-8121/50/2/025401
  [arXiv:1511.08276 [hep-th]].
  %%CITATION = doi:10.1088/1751-8121/50/2/025401;%%
  
\bibitem{cui1}
 X.~Cui and M.~Shifman,
{\em N=(0,2) Deformation of CP(1) Model: Two-dimensional Analog of N=1 Yang-Mills Theory in Four Dimensions,}
  Phys.\ Rev.\ D {\bf 85}, 045004 (2012)
%  doi:10.1103/PhysRevD.85.045004
  [arXiv:1111.6350 [hep-th]].
  %%CITATION = doi:10.1103/PhysRevD.85.045004;%%
  %9 citations counted in INSPIRE as of 23 Feb 2017
  
  \bibitem{Banks}
  T.~Banks and A.~Zaks,
{\em On the Phase Structure of Vector-Like Gauge Theories with Massless Fermions,}
  Nucl.\ Phys.\ B {\bf 196}, 189 (1982).
 % doi:10.1016/0550-3213(82)90035-9
  %%CITATION = doi:10.1016/0550-3213(82)90035-9;%%
  %854 citations counted in INSPIRE as of 01 Apr 2018

\bibitem{cui2}
J.~Chen, X.~Cui, M.~Shifman and A.~Vainshtein,
{\em N=(0,2) deformation of (2, 2) sigma models: Geometric structure, holomorphic anomaly, and exact $\beta$ functions,}
  Phys.\ Rev.\ D {\bf 90}, no. 4, 045014 (2014)
 % doi:10.1103/PhysRevD.90.045014
  [arXiv:1404.4689 [hep-th]].
  %%CITATION = doi:10.1103/PhysRevD.90.045014;%%
  %8 citations counted in INSPIRE as of 23 Feb 2017

\bibitem{shiy}
M. Edalati and D. Tong, {\em Heterotic Vortex Strings}, JHEP {\bf 0705}, 005 (2007)
[hep-th/0703045 [HEP-TH]];
M. Shifman and A. Yung, {\em Heterotic Flux Tubes in N = 2 SQCD with N = 1 Preserving Deformations}, Phys. \ Rev. \ D 
{\bf 77}, 125016 (2008) [Erratum-ibid. D {\bf 79}, 049901 (2009)] [arXiv:0803.0158 [hep-th]];
M. Shifman and A. Yung, {\em Two-Dimensional Sigma Models Related to Non-Abelian Strings in Super-Yang-Mills}, arXiv:1401.7067 [published in {\sl Pomeranchuk 100}, Eds. A. Gorsky and M. Vysotsky, (World Scientific, Singapore, 2014), p. 181]. 

\bibitem{SS}
M.~Shifman and K.~Stepanyantz,
{\em Exact Adler Function in Supersymmetric QCD,}
  Phys.\ Rev.\ Lett.\  {\bf 114}, no. 5, 051601 (2015)
  %doi:10.1103/PhysRevLett.114.051601
  [arXiv:1412.3382 [hep-th]];
  %%CITATION = doi:10.1103/PhysRevLett.114.051601;%%
  %17 citations counted in INSPIRE as of 03 Apr 2018
  {\em Derivation of the exact expression for the D function in N=1 SQCD,}
  Phys.\ Rev.\ D {\bf 91}, 105008 (2015)
  %doi:10.1103/PhysRevD.91.105008
  [arXiv:1502.06655 [hep-th]].
  %%CITATION = doi:10.1103/PhysRevD.91.105008;%%
  %15 citations counted in INSPIRE as of 03 Apr 2018
  
  \bibitem{KKS}
  A.~L.~Kataev, A.~E.~Kazantsev and K.~V.~Stepanyantz,
{\em The Adler $D$ function for ${\cal N}=1$ SQCD Regularized by Higher Covariant Derivatives in the Three-Loop Approximation,}
  Nucl.\ Phys.\ B {\bf 926} (2018) 295
  %doi:10.1016/j.nuclphysb.2017.11.009
  [arXiv:1710.03941 [hep-th]].
  %%CITATION = doi:10.1016/j.nuclphysb.2017.11.009;%%

  \bibitem{KS}
  A.~L.~Kataev and K.~V.~Stepanyantz,
{\em NSVZ Scheme with the Higher Derivative Regularization for $\mathcal{N} =$ 1 SQED,}
  Nucl.\ Phys.\ B {\bf 875} (2013) 459
  %doi:10.1016/j.nuclphysb.2013.07.010
  [arXiv:1305.7094 [hep-th]].
  %%CITATION = doi:10.1016/j.nuclphysb.2013.07.010;%%
  
  
  
\bibitem{gc} 
  V.~A.~Novikov, M.~A.~Shifman, A.~I.~Vainshtein and V.~I.~Zakharov,
{\em Instanton Effects in Supersymmetric Theories,}
  Nucl.\ Phys.\ B {\bf 229}, 407 (1983).
  %doi:10.1016/0550-3213(83)90340-1
  %%CITATION = doi:10.1016/0550-3213(83)90340-1;%%
  %237 citations counted in INSPIRE as of 23 Feb 2017
  
  \bibitem{Novikov} 
  V.~A.~Novikov, M.~A.~Shifman, A.~I.~Vainshtein and V.~I.~Zakharov,
{\em Supersymmetric Instanton Calculus (Gauge Theories with Matter),}
  Nucl.\ Phys.\ B {\bf 260}, 157 (1985);
  %%CITATION = doi:10.1016/0550-3213(85)90316-5;%%
  %201 citations counted in INSPIRE as of 23 Feb 2017
  
  \bibitem{masv}
  
   M.~A.~Shifman and A.~I.~Vainshtein,
{\em On Gluino Condensation in Supersymmetric Gauge Theories. SU(N) and O(N) Groups,}
  Nucl.\ Phys.\ B {\bf 296}, 445 (1988);
  %%CITATION = doi:10.1016/0550-3213(88)90680-3;%%
  %151 citations counted in INSPIRE as of 23 Feb 2017
   A.~Y.~Morozov, M.~A.~Olshanetsky and M.~A.~Shifman,
{\em Gluino Condensate in Supersymmetric Gluodynamics,}
 Nucl.\ Phys.\ B {\bf 304}, 291 (1988).
  %%CITATION = doi:10.1016/0550-3213(88)90628-1;%%
  %29 citations counted in INSPIRE as of 23 Feb 2017
  
   \bibitem{khoze}
   N.~M.~Davies, T.~J.~Hollowood, V.~V.~Khoze and M.~P.~Mattis,
{\em Gluino condensate and magnetic monopoles in supersymmetric gluodynamics,}
  Nucl.\ Phys.\ B {\bf 559}, 123 (1999)
  %doi:10.1016/S0550-3213(99)00434-4
  [hep-th/9905015].
  %%CITATION = doi:10.1016/S0550-3213(99)00434-4;%%
  %167 citations counted in INSPIRE as of 05 Jan 2017
  
\bibitem{Nekrasov} 
  N.~A.~Nekrasov,
{\em Seiberg-Witten prepotential from instanton counting,}
  Adv.\ Theor.\ Math.\ Phys.\  {\bf 7}, no. 5, 831 (2003)
 % doi:10.4310/ATMP.2003.v7.n5.a4
  [hep-th/0206161].
  %%CITATION = doi:10.4310/ATMP.2003.v7.n5.a4;%%
  %883 citations counted in INSPIRE as of 23 Feb 2017

\bibitem{A2}
J.~T.~\L{}opusza\'{n}ski and M.~Sohnius,
%{\em On The Algebra Of Super-Symmetry Transformations,}
Karlsruhe Report Print-74-1269 (unpublished).

\bibitem{A3}
R.~Haag, J.~T.~\L{}opusza\'{n}ski and M.~Sohnius,
%{\em All Possible Generators Of Supersymmetries Of The S Matrix,}
Nucl.\ Phys.\ B {\bf 88}, 257 (1975)
[Reprinted in {\em Supersymmetry},
Ed. S. Ferrara, (North-Holland/World Scientific, 1987) Vol. 1, p.  51].
%%CITATION = NUPHA,B88,257;%%

\bibitem{Olive}
E.~Witten and D.~I.~Olive,
{\em  Supersymmetry Algebras That Include Topological Charges,}
  Phys.\ Lett.\  {\bf 78B}, 97 (1978).
  %doi:10.1016/0370-2693(78)90357-X
  %%CITATION = doi:10.1016/0370-2693(78)90357-X;%%
  %892 citations counted in INSPIRE as of 03 Apr 2018
  
  \bibitem{Shifman} 
  M.~A.~Shifman, A.~I.~Vainshtein and M.~B.~Voloshin,
{\em Anomaly and quantum corrections to solitons in two-dimensional theories with minimal supersymmetry,}
  Phys.\ Rev.\ D {\bf 59}, 045016 (1999)
 % doi:10.1103/PhysRevD.59.045016
  [hep-th/9810068].
  %%CITATION = doi:10.1103/PhysRevD.59.045016;%%
  %90 citations counted in INSPIRE as of 25 Feb 2017
  
  \bibitem{LSV}
  A.~Losev, M.~A.~Shifman and A.~I.~Vainshtein,
{\em Counting supershort supermultiplets,}
  Phys.\ Lett.\ B {\bf 522}, 327 (2001)
  %doi:10.1016/S0370-2693(01)01293-X
  [hep-th/0108153];
  %%CITATION = doi:10.1016/S0370-2693(01)01293-X;%%
  %29 citations counted in INSPIRE as of 26 Feb 2017
 {\em  Single state supermultiplet in (1+1)-dimensions,}
  New J.\ Phys.\  {\bf 4}, 21 (2002)
  %doi:10.1088/1367-2630/4/1/321
  [hep-th/0011027].
  %%CITATION = doi:10.1088/1367-2630/4/1/321;%%
  %31 citations counted in INSPIRE as of 26 Feb 2017

\bibitem{DvSh}
G.~R.~Dvali and M.~A.~Shifman,
%``Domain walls in strongly coupled theories,''
Phys.\ Lett.\ B {\bf 396}, 64 (1997)
(E)\ B {\bf 407}, 452 (1997)
[hep-th/9612128].
%%CITATION = HEP-TH 9612128;%%

\bibitem{Witten} 
  E.~Witten,
{\em Branes and the dynamics of QCD,}
  Nucl.\ Phys.\ B {\bf 507}, 658 (1997)
%  doi:10.1016/S0550-3213(97)00648-2
  [hep-th/9706109].
  %%CITATION = doi:10.1016/S0550-3213(97)00648-2;%%
  %425 citations counted in INSPIRE as of 27 Feb 2017
  %%CITATION = HEP-TH 9909015;%%
 
\bibitem{Achar}
  B.~S.~Acharya and C.~Vafa,
 {\em On domain walls of N=1 supersymmetric Yang-Mills in four-dimensions,}
  hep-th/0103011.
  %%CITATION = HEP-TH/0103011;%%
  %102 citations counted in INSPIRE as of 27 Feb 2017
  
  \bibitem{kovner}
  A.~Kovner, M.~A.~Shifman and A.~V.~Smilga,
{\em Domain walls in supersymmetric Yang-Mills theories,}
  Phys.\ Rev.\ D {\bf 56}, 7978 (1997)
  doi:10.1103/PhysRevD.56.7978
  [hep-th/9706089].
  %%CITATION = doi:10.1103/PhysRevD.56.7978;%%
  %132 citations counted in INSPIRE as of 03 Apr 2018
  
  \bibitem{svritz}
   A.~Ritz, M.~Shifman and A.~Vainshtein,
 {\em Counting domain walls in N=1 superYang-Mills,}
  Phys.\ Rev.\ D {\bf 66}, 065015 (2002)
%  doi:10.1103/PhysRevD.66.065015
  [hep-th/0205083];
  %%CITATION = doi:10.1103/PhysRevD.66.065015;%%
  %24 citations counted in INSPIRE as of 27 Feb 2017
  
    \bibitem{svritz2}
   A.~Ritz, M.~Shifman and A.~Vainshtein,
{\em Enhanced worldvolume supersymmetry and intersecting domain walls in N=1 SQCD,}
  Phys.\ Rev.\ D {\bf 70}, 095003 (2004)
  %doi:10.1103/PhysRevD.70.095003
  [hep-th/0405175].
  %%CITATION = doi:10.1103/PhysRevD.70.095003;%%
  %27 citations counted in INSPIRE as of 03 Apr 2018
  
    \bibitem{ritz}
  A. Ritz and A. Shukla, {\em Domain wall moduli in softly broken SQCD
  at $\theta =\pi$}, to appear.
  
  \bibitem{gai} 
  D.~Gaiotto, A.~Kapustin, Z.~Komargodski and N.~Seiberg,
  {\em Theta, Time Reversal, and Temperature,}
  JHEP {\bf 1705}, 091 (2017)
  %doi:10.1007/JHEP05(2017)091
  [arXiv:1703.00501 [hep-th]].
  %%CITATION = doi:10.1007/JHEP05(2017)091;%%
  %40 citations counted in INSPIRE as of 03 Apr 2018
  
  \bibitem{Losev} 
  A.~Losev and M.~Shifman,
{\em N=2 sigma model with twisted mass and superpotential: Central charges and solitons,}
  Phys.\ Rev.\ D {\bf 68}, 045006 (2003)
 % doi:10.1103/PhysRevD.68.045006
  [hep-th/0304003] (Section 10);
  %%CITATION = doi:10.1103/PhysRevD.68.045006;%%
  %19 citations counted in INSPIRE as of 27 Feb 2017
  M.~Shifman, A.~Vainshtein and R.~Zwicky,
 {\em Central charge anomalies in 2-D sigma models with twisted mass,}
  J.\ Phys.\ A {\bf 39}, 13005 (2006)
  %doi:10.1088/0305-4470/39/41/S13
  [hep-th/0602004].
  %%CITATION = doi:10.1088/0305-4470/39/41/S13;%%
  %31 citations counted in INSPIRE as of 27 Feb 2017

  \bibitem{book}
  M. Shifman and A. Yung,
  {\sl Supersymmetric Solitons}, (Cambridge University Press, 2009). 
 
   \bibitem{used} 
  M. Shifman and A. Yung, {\em Two-Dimensional Sigma Models Related to Non-Abelian Strings in Super-Yang-Mills}, arXiv:1401.7067 [published in {\sl Pomeranchuk 100}, Eds. A. Gorsky and M. Vysotsky, (World Scientific, Singapore, 2014), p. 181]. 
    
    \bibitem{armoni}
{\em From superYang-Mills theory to QCD: Planar equivalence and its implications,}
  [hep-th/0403071], published in {\sl From Fields to Strings: Circumnavigating Theoretical Physics}, Eds. M. Shifman, A. Vainshtein and J. Wheater (World Scientific, Singapore, 2005), Vol. 1, page 353.
  %%CITATION = doi:10.1142/9789812775344_0013;%%
  %102 citations counted in INSPIRE as of 27 Feb 2017
  
  \bibitem{ASV}
  A.~Armoni, M.~Shifman and G.~Veneziano,
{\em Exact results in non-supersymmetric large N orientifold field theories,}
  Nucl.\ Phys.\ B {\bf 667}, 170 (2003)
%  doi:10.1016/S0550-3213(03)00538-8
  [hep-th/0302163];
  %%CITATION = doi:10.1016/S0550-3213(03)00538-8;%%
  %191 citations counted in INSPIRE as of 27 Feb 2017
{\em Refining the proof of planar equivalence,}
  Phys.\ Rev.\ D {\bf 71}, 045015 (2005)
 % doi:10.1103/PhysRevD.71.045015
  [hep-th/0412203].
  %%CITATION = doi:10.1103/PhysRevD.71.045015;%%
  %79 citations counted in INSPIRE as of 27 Feb 2017
    
    \bibitem{ASV2}
    A.~Armoni, M.~Shifman and G.~Veneziano,
{\em QCD quark condensate from SUSY and the orientifold large N expansion,}
  Phys.\ Lett.\ B {\bf 579}, 384 (2004)
 % doi:10.1016/j.physletb.2003.10.094
  [hep-th/0309013];
  %%CITATION = doi:10.1016/j.physletb.2003.10.094;%%
  %61 citations counted in INSPIRE as of 27 Feb 2017
   A.~Armoni, M.~Shifman, G.~Shore and G.~Veneziano,
{\em The quark condensate in multi-flavour QCD: planar equivalence confronting lattice simulations,}
  Phys.\ Lett.\ B {\bf 741}, 184 (2015)
  %doi:10.1016/j.physletb.2014.12.035
  [arXiv:1412.3389 [hep-th]].
  %%CITATION = doi:10.1016/j.physletb.2014.12.035;%%
  %4 citations counted in INSPIRE as of 27 Feb 2017
  
  \bibitem{Cohen}
{\em All you need is N: Baryon spectroscopy in two large N limits,}
  Phys.\ Rev.\ D {\bf 80}, 036002 (2009)
%  doi:10.1103/PhysRevD.80.036002
  [arXiv:0906.2400 [hep-ph]];
  %%CITATION = doi:10.1103/PhysRevD.80.036002;%%
  %22 citations counted in INSPIRE as of 03 Apr 2018
M.~I.~Buchoff, A.~Cherman and T.~D.~Cohen,
 {\em Color Superconductivity at Large N: A New Hope,}
  Phys.\ Rev.\ D {\bf 81}, 125021 (2010)
  %doi:10.1103/PhysRevD.81.125021
  [arXiv:0910.0470 [hep-ph]];
  %%CITATION = doi:10.1103/PhysRevD.81.125021;%%
  %10 citations counted in INSPIRE as of 03 Apr 2018
F.~Buisseret and C.~Semay,
  %``Light baryon masses in different large-$N_c$ limits,''
  Phys.\ Rev.\ D {\bf 82}, 056008 (2010)
  %doi:10.1103/PhysRevD.82.056008
  [arXiv:1006.4729 [hep-ph]];
  %%CITATION = doi:10.1103/PhysRevD.82.056008;%%
  %7 citations counted in INSPIRE as of 03 Apr 2018
  T.~D.~Cohen,
 {\em Nuclear and hadronic physics at large $N_c$,}
  PoS QNP {\bf 2012}, 022 (2012).
  %%CITATION = POSCI,QNP2012,022;%%
  
  \bibitem{Cohen2} 
  T.~D.~Cohen and R.~F.~Lebed,
  {\em Tetraquarks with exotic flavor quantum numbers at large $N_c$ in QCD(AS),}
  Phys.\ Rev.\ D {\bf 89}, no. 5, 054018 (2014)
%doi:10.1103/PhysRevD.89.054018
  [arXiv:1401.1815 [hep-ph]].
  %%CITATION = doi:10.1103/PhysRevD.89.054018;%%
  %17 citations counted in INSPIRE as of 03 Apr 2018
  
  \bibitem{ANO}
A.~A.~Abrikosov,
{\em On the Magnetic properties of superconductors of the second kind,"}
  Sov.\ Phys.\ JETP {\bf 5}, 1174 (1957);
  %%CITATION = SPHJA,5,1174;%%
  %802 citations counted in INSPIRE as of 27 Feb 2017.~Abrikosov, Sov.~Phys. JETP {\bf32} 1442  (1957);
H.~B.~Nielsen and P.~Olesen,
{\em Vortex Line Models for Dual Strings,}
  Nucl.\ Phys.\ B {\bf 61}, 45 (1973).
 % doi:10.1016/0550-3213(73)90350-7
  %%CITATION = doi:10.1016/0550-3213(73)90350-7;%%
  %2253 citations counted in INSPIRE as of 27 Feb 2017

\bibitem{ayung}
A.~Hanany and D.~Tong,
{\em Vortices, instantons and branes,}
  JHEP {\bf 0307}, 037 (2003)
%  doi:10.1088/1126-6708/2003/07/037
  [hep-th/0306150];
  %%CITATION = doi:10.1088/1126-6708/2003/07/037;%%
  %366 citations counted in INSPIRE as of 28 Feb 2017
  {\em Vortex strings and four-dimensional gauge dynamics,}
  JHEP {\bf 0404}, 066 (2004)
 % doi:10.1088/1126-6708/2004/04/066
  [hep-th/0403158];
  %%CITATION = doi:10.1088/1126-6708/2004/04/066;%%
  %252 citations counted in INSPIRE as of 28 Feb 2017
 R.~Auzzi, S.~Bolognesi, J.~Evslin, K.~Konishi and A.~Yung,
{\em Non-Abelian superconductors: Vortices and confinement in N=2 SQCD,}
  Nucl.\ Phys.\ B {\bf 673}, 187 (2003)
%  doi:10.1016/j.nuclphysb.2003.09.029
  [hep-th/0307287];
  %%CITATION = doi:10.1016/j.nuclphysb.2003.09.029;%%
  %315 citations counted in INSPIRE as of 20 Dec 2016
 M.~Shifman and A.~Yung,
{\em Non-Abelian string junctions as confined monopoles,}
  Phys.\ Rev.\ D {\bf 70}, 045004 (2004)
 % doi:10.1103/PhysRevD.70.045004
  [hep-th/0403149].
  %%CITATION = doi:10.1103/PhysRevD.70.045004;%%
  %240 citations counted in INSPIRE as of 20 Dec 2016
  
  \bibitem{rn1}
  M.~Shifman and A.~Yung,
{\em Quantum Deformation of the Effective Theory on Non-Abelian string and 2D-4D correspondence,}
  Phys.\ Rev.\ D {\bf 89}, no. 6, 065035 (2014)
%  doi:10.1103/PhysRevD.89.065035
  [arXiv:1401.1455 [hep-th]].
  %%CITATION = doi:10.1103/PhysRevD.89.065035;%%
  %5 citations counted in INSPIRE as of 01 Mar 2017
  
  \bibitem{Dorey}
  N.~Dorey,
{\em The BPS spectra of two-dimensional supersymmetric gauge theories with twisted mass terms,}
  JHEP {\bf 9811}, 005 (1998)
%  doi:10.1088/1126-6708/1998/11/005
  [hep-th/9806056].
  %%CITATION = doi:10.1088/1126-6708/1998/11/005;%%
  %128 citations counted in INSPIRE as of 28 Feb 2017
  
  \bibitem{het1}
   M.~Edalati and D.~Tong,
 {\em Heterotic Vortex Strings,}
  JHEP {\bf 0705}, 005 (2007)
%  doi:10.1088/1126-6708/2007/05/005
  [hep-th/0703045 [HEP-TH]].
  %%CITATION = doi:10.1088/1126-6708/2007/05/005;%%
  %55 citations counted in INSPIRE as of 28 Feb 2017
  
    \bibitem{het2}
   M.~Shifman and A.~Yung,
  {\em Heterotic Flux Tubes in N=2 SQCD with N=1 Preserving Deformations,}
  Phys.\ Rev.\ D {\bf 77}, 125016 (2008)
  Erratum: [Phys.\ Rev.\ D {\bf 79}, 049901 (2009)]
  %doi:10.1103/PhysRevD.79.049901, 10.1103/PhysRevD.77.125016
  [arXiv:0803.0158 [hep-th]].
  %%CITATION = doi:10.1103/PhysRevD.79.049901, 10.1103/PhysRevD.77.125016;%%
  %42 citations counted in INSPIRE as of 28 Feb 2017
  
  \bibitem{het3}
  M.~Shifman and A.~Yung,
  {\em Large-N Solution of the Heterotic N=(0,2) Two-Dimensional CP(N-1) Model,}
  Phys.\ Rev.\ D {\bf 77}, 125017 (2008),
  Erratum: [Phys.\ Rev.\ D {\bf 81}, 089906 (2010)];
%  doi:10.1103/PhysRevD.77.125017, 10.1103/PhysRevD.81.089906
  [arXiv:0803.0698 [hep-th]];
  %%CITATION = doi:10.1103/PhysRevD.77.125017, 10.1103/PhysRevD.81.089906;%%
  %37 citations counted in INSPIRE as of 28 Feb 2017
  P.~A.~Bolokhov, M.~Shifman and A.~Yung,
 {\em Large-$N$ Solution of the Heterotic $CP(N-1)$ Model with Twisted Masses,}
  Phys.\ Rev.\ D {\bf 82}, no. 2, 025011 (2010)
  Erratum: [Phys.\ Rev.\ D {\bf 89}, no. 2, 029904 (2014)]
  %doi:10.1103/PhysRevD.89.029904, 10.1103/PhysRevD.82.025011
  [arXiv:1001.1757 [hep-th]].
  %%CITATION = doi:10.1103/PhysRevD.89.029904, 10.1103/PhysRevD.82.025011;%%
  %11 citations counted in INSPIRE as of 28 Feb 2017  
  
  \bibitem{PS}
  J.~Polchinski and A.~Strominger,
{\em Effective string theory,}
  Phys.\ Rev.\ Lett.\  {\bf 67}, 1681 (1991).
  %doi:10.1103/PhysRevLett.67.1681
  %%CITATION = doi:10.1103/PhysRevLett.67.1681;%%
  %180 citations counted in INSPIRE as of 28 Feb 2017
  
  \bibitem{cs1}
  M.~Shifman and A.~Yung,
{\em Critical String from Non-Abelian Vortex in Four Dimensions,}
  Phys.\ Lett.\ B {\bf 750}, 416 (2015)
%  doi:10.1016/j.physletb.2015.09.045
  [arXiv:1502.00683 [hep-th]].
  %%CITATION = doi:10.1016/j.physletb.2015.09.045;%%
  %6 citations counted in INSPIRE as of 28 Feb 2017
  
  \bibitem{cs2}
  P.~Koroteev, M.~Shifman and A.~Yung,
{\em Studying Critical String Emerging from Non-Abelian Vortex in Four Dimensions,}
  Phys.\ Lett.\ B {\bf 759}, 154 (2016)
%  doi:10.1016/j.physletb.2016.05.075
  [arXiv:1605.01472 [hep-th]];
  %%CITATION = doi:10.1016/j.physletb.2016.05.075;%%
  %4 citations counted in INSPIRE as of 28 Feb 2017
{\em Non-Abelian Vortex in Four Dimensions as a Critical String on a Conifold,}
  Phys.\ Rev.\ D {\bf 94}, no. 6, 065002 (2016)
 % doi:10.1103/PhysRevD.94.065002
  [arXiv:1605.08433 [hep-th]];
  %%CITATION = doi:10.1103/PhysRevD.94.065002;%%
  %3 citations counted in INSPIRE as of 28 Feb 2017
{\em Non-Abelian Vortex in Four Dimensions as a Critical Superstring,}
JETP Letters, {\bf 105}, no.1, 60
  %doi:10.1134/S0021364017010040
  arXiv:1611.03111 [hep-th].
  %%CITATION = doi:10.1134/S0021364017010040;%%
  
\bibitem{sl}
  M.~Shifman and A.~Yung,
{\em Non-Abelian semilocal strings in N=2 supersymmetric QCD,}
  Phys.\ Rev.\ D {\bf 73}, 125012 (2006)
  %doi:10.1103/PhysRevD.73.125012
  [hep-th/0603134].
  %%CITATION = doi:10.1103/PhysRevD.73.125012;%%
  %89 citations counted in INSPIRE as of 28 Feb 2017
  
  \bibitem{SVY}
   M.~Shifman, W.~Vinci and A.~Yung,
  {\em Effective World-Sheet Theory for Non-Abelian Semilocal Strings in N = 2 Supersymmetric QCD,}
  Phys.\ Rev.\ D {\bf 83}, 125017 (2011)
%  doi:10.1103/PhysRevD.83.125017
  [arXiv:1104.2077 [hep-th]].
  %%CITATION = doi:10.1103/PhysRevD.83.125017;%%
  %19 citations counted in INSPIRE as of 01 Mar 2017
  
  \bibitem{ArgPlessShapiro}
P.~Argyres, M.~R.~Plesser and  A.~Shapere,
{\em The Coulomb Phase of N=2 Supersymmetric QCD}
Phys.\ Rev. \ Lett. {\bf 75}, 1699 (1995).
[hep-th/9505100].
  
  \bibitem{APS}
P.~Argyres, M.~Plesser and N.~Seiberg,
{\em The Moduli Space of ${\mathcal N}=2$  SUSY QCD and Duality in
${\mathcal N}=1$  SUSY QCD,}
Nucl. Phys. {\bf B471}, 159  (1996).
  
  \bibitem{Kom}
 E.~Gerchkovitz and A.~Karasik,
{\em New Vortex-String Worldsheet Theories from Supersymmetric Localization,}
  arXiv:1711.03561 [hep-th].
  %%CITATION = ARXIV:1711.03561;%%
  %1 citations counted in INSPIRE as of 19 Apr 2018
  
\bibitem{NVafa}
A.~Neitzke and  C.~Vafa,
{\em Topological strings and their physical applications},
arXiv:hep-th/0410178.

  \bibitem{GukVafaWitt}
S.~Gukov, C.~Vafa and E.~Witten
{\em CFT's from Calabi-Yau four folds, }
Nucl. \ Phys. \ {\bf B584},  69 (2000)
[arXiv:hep-th/0410178].

\bibitem{Candel}
P.~Candelas and X.~C.~ de la Ossa,
{\em Comments on conifolds,}
Nucl. \ Phys. \ {\bf B342}, 246 (1990).

\bibitem{Zayas}
L.~A.~Pando Zayas and  A.~A.~Tseytlin,
{\em 3-branes on resolved conifold,}
 JHEP \ {\bf 0011},  028 (2000)
[arXiv:hep-th/0010088].

\bibitem{Klebanov}
I.~R.~Klebanov and  A.~Murugan,
{\em Gauge/Gravity Duality and Warped Resolved Conifold,}
JHEP \ {\bf 0703}, 042 (2007)
[arXiv:hep-th/0701064].

\bibitem{GivKut}
A.~Giveon and D.~Kutasov,
{\em Little string theory in a double scaling limit},
JHEP {\bf 9910}  034, (1999)
[arXiv: hep-th/9909110]

\bibitem{GVafa}
D.~Ghoshal and  C.~Vafa, 
{\em c = 1 string as the topological theory of the conifold},
 Nucl. \ Phys.\  {\bf B453} 121, (1995)
[arXiv: hep-th/9506122 ]
  
%\bibitem{GivKutP}
%A.~Giveon, D.~Kutasov and O.~Pelc,
%{\em Holography for noncritical superstrings},
%JHEP {\bf 9910}  035, (1999)
%[arXiv: hep-th/9907178].
%
%\bibitem{ABKS}
%O.~Aharony, M.~Berkooz, D.~Kutasov and  N.~Seiberg 
%{\em Linear dilatons, NS five-branes and holography},
%JHEP {\bf 9810} 004,  (1998)
%[arXiv:hep-th/9808149].
%
%\bibitem{Wbh}
%E.~Witten,
%{\em String theory and black holes,}
%Phys.\ Rev.\  D {\bf 44}, 314 (1991)
%
%\bibitem{MukVafa}
%S.~Mukhi and C.~Vafa,
%{\em Two-dimensional black hole as a topological coset model of c = 1 string theory},
%Nucl. \ Phys.\ {\bf B407}  667, (1993)
%[arXiv: hep-th/9301083].
%
%\bibitem{OoguriVafa95}
%H.~Ooguri and C.~Vafa,
%{\em Two-Dimensional Black Hole and Singularities of CY Manifolds,}
%Nucl. \ Phys. \ {\bf B463} 55, (1996)
[arXiv:hep-th/9511164].

\bibitem{sylbo}
 M.~Shifman and A.~Yung,
{\em Critical Non-Abelian Vortex in Four Dimensions and Little String Theory,}
  Phys.\ Rev.\ D {\bf 96}, no. 4, 046009 (2017)
  %doi:10.1103/PhysRevD.96.046009
  [arXiv:1704.00825 [hep-th]].
  %%CITATION = doi:10.1103/PhysRevD.96.046009;%%
  %2 citations counted in INSPIRE as of 03 Apr 2018
  
  \bibitem{tbp}
 M.~Shifman and A.~Yung, to be published.
  
\bibitem{Fey}

Richard Feynman, {\sl  The Character of Physical Law},
( MIT Press, 1965).

  

  

 }
 
 
 
\end{thebibliography}
\end{document}